\documentclass[aps,twocolumn,floatfix,showpacs,amsmath,amsfonts]{revtex4}

\usepackage{graphicx}
\usepackage{bm}
\usepackage{units}
\usepackage{color}
\usepackage{upgreek}

\begin{document}

\title{GOE-GUE-Poisson transitions in the nearest neighbor spacing distribution of magnetoexcitons}

\author{Frank Schweiner}
\author{J\"org Main}
\author{G\"unter Wunner}
\affiliation{Institut f\"ur Theoretische Physik 1, Universit\"at Stuttgart,
  70550 Stuttgart, Germany}
\date{\today}

\begin{abstract}
Recent investigations on the Hamiltonian of excitons by F.~Schweiner~\emph{et al.}
[Phys. Rev. Lett.~\textbf{118}, 046401 (2017)]
revealed that the combined presence of a cubic band structure and external fields
breaks all antiunitary symmetries.
The nearest neighbor spacing distribution of magnetoexcitons 
can exhibit Poissonian statistics, the statistics of a Gaussian orthogonal ensemble (GOE)
or a Gaussian unitary ensemble (GUE) depending on the system parameters.
Hence, magnetoexcitons are an
ideal system to investigate the transitions between these statistics.
Here we investigate the transitions
between GOE and GUE statistics and between Poissonian and GUE statistics
by changing the angle of the magnetic field with respect to
the crystal lattice and by changing the scaled energy
known from the hydrogen atom in external fields.
Comparing our results with analytical formulae for these transitions
derived with random matrix theory, we obtain a very good agreement and
thus confirm the Wigner surmise for the exciton system.
\end{abstract}


\maketitle

\section{Introduction}

Ever since the Bohigas-Giannoni-Schmit conjecture~\cite{QC_1}, which
stated that these quantum systems can be described by random
matrix theory~\cite{QSC_29,QSC_30}, it has been shown that
irregular classical behavior manifests itself in statistical quantities
of the corresponding quantum system~\cite{GUE5}.
In random matrix theory the Hamiltonian of a 
system is replaced by a random matrix with appropriate symmetries
to study the statistical properties of its eigenvalue spectrum~\cite{GUE4};
so only universal quantities of a system are considered
and detailed dynamical properties are irrelevant.
Even though Hamiltonians of dynamical systems are not 
random in most cases, it is already understood 
that spectral fluctuations for 
nonrandom and random Hamiltonians are equivalent~\cite{GUE3,QSC,GUE3_3}.

All systems with a Hamiltonian leading to global chaos in the classical dynamics
can be assigned to one of three universality classes: the orthogonal, the unitary or
the symplectic universality class~\cite{QSC}.
To which of these universality classes a given system
belongs is determined by the remaining symmetries in the system.
Most of the physical systems still have time-reversal 
or at least one remaining antiunitary symmetry
and thus show the statistics of a Gaussion orthogonal ensemble (GOE).
Some examples of these systems are 
nuclei in external magnetic fields~\cite{QC_5,QSC_11,QSC_12,QSC_13},
microwave billards~\cite{QSC_15,QSC_16,QSC_17}, molecular spectra~\cite{QSC_18},
impurities~\cite{QC_3}, and quantum wells~\cite{QC_4}.
Atoms in constant external fields, in particular, 
are among the most important physical systems belonging to
the orthogonal universality class~\cite{QSC_19,QC_2,GUE1}.
They are ideal systems to investigate 
the emergence of quantum chaos both in high-precision
experimental measurements and precise quantal calculations,
possible because of the availability of the analytically known Hamiltonian
(see Refs.~\cite{GUE9,GUE5} and further references therein). 
Hence, they are a perfectly suitable physical system to study the transition from
the Poissonian level statistics, which describes the classically integrable case in the absence of the fields~\cite{QSC,GUE3_6}, 
to GOE statistics~\cite{GUE1}, where the breaking of symmetries due to the external fields 
leads to a correlation of levels and hence to a strong
suppression of crossings~\cite{QSC}.

As regards the other universality classes, examples are much rarer
since systems \emph{without any antiunitary symmetry} [Gaussian unitary ensemble (GUE)] or
systems with \emph{time-reversal invariance possessing Kramer's degeneracy
but no geometric symmetry at all} [Gaussian symplectic ensemble (GSE)] have to be found~\cite{QSC}.
Until now GUE statistics was observable in rather exotic systems
such as microwave cavities with ferrite strips~\cite{QSC_27},
atoms in a static electric field and a resonant microwave field of elliptical polarization~\cite{GUE8},
a kicked rotor or a kicked top~\cite{GUE3,GUE4_16,QSC_K7_36},
the metal-insulator transition in the Anderson model of disordered systems~\cite{GUE4_21},
which can be compared to the Brownian motion model~\cite{GUE4_22},
or for billards in microwave resonators~\cite{QSC_26},
and in graphene quantum dots~\cite{QC_9}.
Since random matrix theory has already been 
extended to describe also transitions between the different
statistics with analytical functions~\cite{GUE4}, 
it is highly desirable to study these transitions
theoretically and experimentally. However, due to the
small number of physical systems showing GUE statistics,
there are only few examples, where transitions
from Poissonian to GUE statistics or from GOE to GUE statistics in dependence of a 
parameter of the system could be studied~\cite{GUE7,GUE3,GUE4_16,QSC_K7_36,GUE4_21}.
Often only mathematical models with specifically designed Hamiltonians
are introduced to investigate these transitions~\cite{GUE4}.

In this paper we will investigate these transitions
in magnetoexcitons.
Excitons are the fundamental optical excitations in the
visible or ultraviolet spectrum of a semiconductor
and consist of an electron in the conduction
band and a positively charged hole in the valence band.
As the interaction between both quasi particles can be described by
a screened Coulomb interaction, excitons are often regarded as the
hydrogen analog of the solid state.
Only three years ago T.~Kazimierczuk~\emph{et~al}~\cite{GRE} 
observed in a remarkable 
high-resolution absorption experiment
an almost perfect hydrogen-like absorption series
for the yellow exciton in cuprous oxide $\left(\mathrm{Cu_{2}O}\right)$ 
up to a principal quantum number of $n=25$. 
This experiment has opened the field of research of giant Rydberg excitons,
and has stimulated a large number of experimental and theoretical 
investigations~\cite{GRE,QC,QC2,175,75,76,50,28,80,100,125,78,79,150,74,77,200}.

Very recently, we have shown that the 
Hamiltonian of magnetoexcitons in cubic semiconductors
breaks all antiunitary symmetries~\cite{175}.
This is the first evidence for a
spatially homogeneous system breaking all antiunitary symmetries.

Since in many cases excitons are treated theoretically via a hydrogen-like Hamiltonian,
the appearance of GUE statistics seems surprising
as the hydrogen atom in external fields still shows one antiunitary symmetry.
However, it is well known that the hydrogen-like model of excitons
is often too simple to account for the huge number of effects in the solid 
(see, e.g., Refs.~\cite{SO,SST,TOE,7,1,28,100,75} and further references therein).
M.~A{\ss}mann~\emph{et al.}~\cite{QC,QC2} attributed the appearance of
GUE statistics in a recent experiment with magnetoexcitons in 
$\mathrm{Cu_{2}O}$ to the interaction of excitons with phonons.

However, we have shown that it is indispensable to
account for the complete valence band structure 
to describe the spectra of excitons in magnetic fields in a theoretically correct way~\cite{125}.
Without the complete band structure the striking experimental finding
of a dependence of the magnetoexciton spectra on the direction
of the external magnetic field cannot be explained.
It is indeed
the simultaneous presence of the cubic band structure and external fields
which breaks all antiunitary symmetries and leads to GUE statistics~\cite{175}.

In this paper we investigate the symmetry breaking for excitons
in semiconductors with a cubic band structure
in dependence on system parameters such as the strength and the angle 
of the magnetic field or the scaled energy~\cite{GUE5,GUE5_23}.
Since the eigenvalue spectrum of the magnetoexciton Hamiltonian 
shows Poissonian, GOE \emph{or} GUE statistics depending on these parameters,
it is an ideal system to investigate the transitions between 
GOE and GUE or Poisson and GUE statistics.
To the best of our knowledge, there are only two more systems where both
transitions have been studied, i.e., the kicked top~\cite{QSC_K7_36} 
and the Anderson model~\cite{GUE4_21}. 
However, while the kicked top is a time-dependent system, which has to be treated
within Floquet theory~\cite{GUE3,QSC_K7_36},
the Anderson model is rather a model system
for a $d$-dimensional disordered lattice,
where parameters such as the disorder and the hopping rate
need to be adjusted~\cite{GUE4_21}.
Magnetoexcitons are a more realistic
physical system allowing for a systematic investigation of transitions
between different statistics. In particular, the 
parameters describing these transitions can be easily adjusted in experiments.
Comparing our results with analytical functions from 
random matrix theory describing the transitions between the statistics~\cite{GUE3,GUE4},
we confirm the so-called Wigner surmise~\cite{GUE4_8}, which states that the NNS
of large random matrices can be approximated by the NNS of
$2\times 2$ matrices of the same universality class~\cite{GUE4}.


The paper is organized as follows:
In Sec.~\ref{sec:Theory} we present the Hamiltonian
of excitons in cubic semiconductors in an external magnetic
field and introduce a complete basis to solve the corresponding Schr\"odinger
equation. The methods of solving the Schr\"odinger equation
for fixed values of the external field strenghts or for a constant scaled energy
are discussed in Secs.~\ref{sub:constmagnfield} and~\ref{sec:scaled-energy}, respectively.
Having shown analytically that the presence of the cubic band structure
and external fields breaks all antiunitary symmetries 
in Sec.~\ref{sec:analytical}, we investigate the eigenvalue spectrum and the level
spacing statistics numerically
At first, we demonstrate the appearance of GOE or GUE statistics
for specific directions of an external magnetic field in Sec.~\ref{sub:appearance}.
The transitions between different level spacing statistics are then
investigated in Secs.~\ref{sub:GOE-GUE} and~\ref{sub:P-GUE}.
Finally, we give a short summary and outlook in Sec.~\ref{sec:Summary}.

\section{Hamiltonian and complete basis\label{sec:Theory}}

In this section we briefly discuss the Hamiltonian of
excitons in direct semiconductors with a
cubic valence band structure and show how to
solve the corresponding Schr\"odinger equation
in a complete basis.
For more details see Refs.~\cite{100,125}
and further references therein.

When neglecting external fields at first, the Hamiltonian
of excitons in direct semiconductors is given by~\cite{17_17}
\begin{equation}
H=V\left(\boldsymbol{r}_{e}-\boldsymbol{r}_{h}\right)+H_{\mathrm{e}}\left(\boldsymbol{p}_{\mathrm{e}}\right)+H_{\mathrm{h}}\left(\boldsymbol{p}_{\mathrm{\mathrm{h}}}\right).\label{eq:Hpeph}
\end{equation}
The Coulomb interaction between the electron (e) and the hole (h)
is screened by the dielectric constant~$\varepsilon$:
\begin{equation}
V\left(\boldsymbol{r}_{e}-\boldsymbol{r}_{h}\right)=-\frac{e^{2}}{4\pi\varepsilon_{0}\varepsilon}\frac{1}{\left|\boldsymbol{r}_{e}-\boldsymbol{r}_{h}\right|}.
\end{equation}

Since the conduction band is often parabolic,
the kinetic energy of the electron is similar to that of
a free particle
\begin{equation}
H_{\mathrm{e}}\left(\boldsymbol{p}_{\mathrm{e}}\right)=\frac{\boldsymbol{p}_{\mathrm{e}}^{2}}{2m_{\mathrm{e}}}.
\end{equation}
However, the effective mass $m_{\mathrm{e}}$ of the electron in the semiconductor
has to be used instead of the free electron mass $m_0$.
As regards the valence bands, the situation is more complicated.
In general, the uppermost valence band is threefold degenerate at the 
center of the Brillouin zone or the $\Gamma$ point
and the kinetic energy of a hole within these valence bands is given
by~\cite{80,100}
\begin{eqnarray}
H_{\mathrm{h}}\left(\boldsymbol{p}_{\mathrm{h}}\right) & = & 
\left(1/2\hbar^{2}m_{0}\right)\left\{ \hbar^{2}\left(\gamma_{1}+4\gamma_{2}\right)\boldsymbol{p}_{\mathrm{h}}^{2}\right.\phantom{\frac{1}{1}}\nonumber \\
 & - & 6\gamma_{2}\left(p_{\mathrm{h}1}^{2}\boldsymbol{I}_{1}^{2}+\mathrm{c.p.}\right)\nonumber \\
 & - & 12\gamma_{3}\left(\left\{ p_{\mathrm{h}1},p_{\mathrm{h}2}\right\} \left\{ \boldsymbol{I}_{1},\boldsymbol{I}_{2}\right\} +\mathrm{c.p.}\right)\phantom{\frac{1}{1}}\label{eq:Hh}
\end{eqnarray}
with $\boldsymbol{p}=\left(p_1,\,p_2,\,p_3\right)$, $\left\{ a,b\right\} =\frac{1}{2}\left(ab+ba\right)$ and
c.p.~denoting cyclic permutation.
The three Luttinger parameters $\gamma_{i}$
describe the behavior and the 
anisotropic effective mass of the hole. 
The matrices $\boldsymbol{I}_j$
denote the three
spin matrices of the quasispin $I=1$
which describes the threefold degenerate valence band~\cite{25}.
The components of these matrices $\boldsymbol{I}_{i}$ read~\cite{25,100}
\begin{equation}
I_{i,\,jk}=-i\hbar\varepsilon_{ijk}\label{eq:I}
\end{equation}
with the Levi-Civita symbol $\varepsilon_{ijk}$.

Note that the expression for $H_{\mathrm{h}}\left(\boldsymbol{p}_{\mathrm{h}}\right)$
can be separated in two parts having spherical and cubic symmetry, respectively~\cite{7_11}.
The coefficients $\mu'$ and $\delta'$
of these parts can be expressed in terms of the three Luttinger parameters:
$\mu'=\left(6\gamma_{3}+4\gamma_{2}\right)/5\gamma'_{1}$ and
$\delta'=\left(\gamma_{3}-\gamma_{2}\right)/\gamma'_{1}$
with $\gamma'_{1}=\gamma_{1}+m_{0}/m_{\mathrm{e}}$~\cite{7_11,7,100}.
The spin-orbit coupling $H_{\mathrm{so}}$, which generally enters the kinetic energy
of the hole~(\ref{eq:Hh}), is neglected here since it 
is spherically symmetric and therefore does not affect the symmetry properties
of the exciton Hamiltonian.


%
%
%

When applying external fields, the
corresponding Hamiltonian is obtained via the minimal substitution. 
After introducing relative and center of mass coordinates~\cite{90,91}
and setting the position and momentum of the 
center of mass to zero, the complete Hamiltonian
of the relative motion reads~\cite{34,33,39,TOE,44,90,91}
\begin{eqnarray}
H & = & V\left(\boldsymbol{r}\right)+e\Phi\left(\boldsymbol{r}\right)\nonumber \\
 & + & H_{\mathrm{e}}\left(\boldsymbol{p}+e\boldsymbol{A}\left(\boldsymbol{r}\right)\right)+H_{\mathrm{h}}\left(-\boldsymbol{p}+e\boldsymbol{A}\left(\boldsymbol{r}\right)\right)
 \label{eq:H}
\end{eqnarray}
with the relative coordinate $\boldsymbol{r}=\boldsymbol{r}_{\mathrm{e}}-\boldsymbol{r}_{\mathrm{h}}$
and the relative momentum $\boldsymbol{p}=\left(\boldsymbol{p}_{\mathrm{e}}-\boldsymbol{p}_{\mathrm{h}}\right)/2$
of electron and hole.
We use the vector potential $\boldsymbol{A}=\left(\boldsymbol{B}\times\boldsymbol{r}\right)/2$
of a constant magnetic field $\boldsymbol{B}$ and the electrostatic potential
$\Phi\left(\boldsymbol{r}\right)=-\boldsymbol{F}\cdot\boldsymbol{r}$ of a constant electric field $\boldsymbol{F}$.

As we will show in Sec.~\ref{sec:analytical}, the symmetry
breaking in the system depends on the orientation
of the fields with respect to the crystal lattice.
We will denote the orientation of $\boldsymbol{B}$ and $\boldsymbol{F}$
in spherical coordinates via
\begin{equation}
\boldsymbol{B}\left(\varphi,\,\vartheta\right)=B\left(\begin{array}{c}
\cos\varphi\sin\vartheta,\\
\sin\varphi\sin\vartheta\\
\cos\vartheta
\end{array}\right)\label{eq:spherical_coord}
\end{equation}
and similar for $\boldsymbol{F}$ in what follows.


Before we solve the Schr\"odinger equation corresponding to the Hamiltonian~(\ref{eq:H}),
we rotate the coordinate system to make
the quantization axis coincide with the direction of the magnetic field (see Appendix~\ref{sub:Hamiltonianrm})
and then express the Hamiltonian~(\ref{eq:H}) in terms of irreducible tensors~\cite{ED,7_11,44}.
We can then calculate a matrix representation of
the Schr\"odinger equation using a complete basis. 

Note that the Hamiltonian~(\ref{eq:H}) is a model system for magnetoexcitons
since we neglect the spin-orbit coupling between the quasi spin $I$ and the hole spin
$S_{\mathrm{h}}$, which appears, e.g., in $\mathrm{Cu_{2}O}$~\cite{100,125}.
Furthermore, we neglect an additional term in Eq.~(\ref{eq:H}), which describes the energy
of the electron and hole spin in the magnetic field
but is invariant under the 
symmetry operations considered below.
Therefore, we can disregard these spins in our basis.
As regards the angular momentum part of the basis,
we have to consider that the Hamiltonian~(\ref{eq:H})
couples the angular momentum $L$ of the exciton and the
quasi spin $I$. Hence, we introduce the total momentum $G=L+I$ with the
$z$ component $M_G$. 
For the radial
part of the exciton wave function we use the Coulomb-Sturmian functions of Ref.~\cite{S1}
\begin{equation}
U_{NL}\left(r\right)=N_{NL}\left(2\rho\right)^{L}e^{-\rho}L_{N}^{2L+1}\left(2\rho\right)\label{eq:U}
\end{equation}
with $\rho=r/\alpha$, a normalization factor $N_{NL}$,
the associated Laguerre polynomials $L_{n}^{m}\left(x\right)$ and
an arbitrary scaling parameter $\alpha$. 
Note that we use the radial quantum number
$N$, which is related to the principal quantum number $n$ via $n=N+L+1$.
Finally, we make the following ansatz for the exciton wave function
\begin{subequations}
\begin{eqnarray}
\left|\Psi\right\rangle  & = & \sum_{NLGM_G}c_{NLGM_G}\left|\Pi\right\rangle,\\
\nonumber \\
\left|\Pi\right\rangle  & = & \left|N,\,L,\,I,\,G,\,M_G\right\rangle,\label{eq:basis}
\end{eqnarray}\label{eq:ansatz}%
\end{subequations}
with complex coefficients $c$. 

The Schr\"odinger equation can now be solved for fixed values
of the external field strengths or for a fixed value of the scaled energy
known from atoms in external fields~\cite{GUE5_23}.
Both methods will be presented in the following

\subsection{Constant field strengths~\label{sub:constmagnfield}}

Inserting the ansatz~(\ref{eq:ansatz}) in the Schr\"odinger
equation $H\Psi=E\Psi$ 
yields a matrix representation 
of the Schr\"odinger equation of the form~\cite{175}
\begin{equation}
\boldsymbol{D}\boldsymbol{c}=E\boldsymbol{M}\boldsymbol{c},\label{eq:gev}
\end{equation}
where the external field strengths are assumed to be constant.
The vector $\boldsymbol{c}$ contains the coefficients of the expansion~(\ref{eq:ansatz}).
Since the functions $U_{NL}\left(r\right)$
actually depend on the coordinate $\rho=r/\alpha$, we substitute
$r\rightarrow\rho\alpha$ in the Hamiltonian~(\ref{eq:H}) and multiply
the corresponding Schr\"odinger equation by $\alpha^{2}$.
All matrix elements which enter the hermitian matrices $\boldsymbol{D}$ and
$\boldsymbol{M}$ can be calculated similarly to the 
matrix elements given in Refs.~\cite{100,125}.
The generalized eigenvalue problem~(\ref{eq:gev})
is finally solved using an appropriate LAPACK routine~\cite{Lapack}.

Since in numerical calculations the basis cannot be infinitely large, 
the values of the quantum numbers
are chosen in the following way: For each value of $n=N+L+1\leq n_{\mathrm{max}}$ we use
\begin{eqnarray}
L & = & 0,\,\ldots,\, n-1,\nonumber \\
G & = & \left|L-1\right|,\,\ldots,\,\min\left(L+1,\, G_{\mathrm{max}}\right),\\
M_{G} & = & -G,\,\ldots,\, G\nonumber.
\end{eqnarray}
The values $G_{\mathrm{max}}$ and $n_{\mathrm{max}}$ are chosen
appropriately large so that as many eigenvalues as possible converge. Additionally,
we can use the scaling parameter $\alpha$ to enhance convergence.
In particular, if the eigenvalues of excitonic states with principal
quantum number $n$ are to be be calculated, we can set 
$\alpha=n\gamma_1'\varepsilon a_0$ according to Ref.~\cite{S1}, where
$a_0$ denotes the Bohr radius. 

Note that without an external electric field, 
parity is a good quantum number and
the operators in the Schr\"odinger equation
couple only basis states with even \emph{or} with odd
values of $L$. 
In this case we consider only basis states with
odd values of $L$ as these exciton states can be observed in
parity-forbidden semiconductors~\cite{100,28,74}.

\subsection{Constant scaled energy\label{sec:scaled-energy}}

Besides solving the Schr\"odinger equation or the 
generalized eigenvalue problem~(\ref{eq:gev}) for fixed values
of the external field strength, it is also possible
to use the concept of scaled energy~\cite{GUE5_23}.
In classical mechanics the Hamiltonian of a hydrogen atom in external fields
possesses a scaling property which allows
reducing the three parameters energy $E$, magnetic field $B$ and
electric field $F$ to two parameters~\cite{MA_24,MA_56}.
The corresponding transformation reads
\begin{alignat}{4}
\hat{\boldsymbol{r}}= &\; \gamma^{2/3}\boldsymbol{r},\qquad && \hat{\boldsymbol{p}} &&=\; \gamma^{-1/3}\boldsymbol{p},\nonumber \\
\hat{\boldsymbol{F}}= &\; \gamma^{-4/3}\boldsymbol{F},\qquad && \hat{E} &&=\; \gamma^{-2/3}E,
\end{alignat}
with $\gamma=B/B_0$ and $B_0=2.3505\times 10^5\,\mathrm{T}$~\cite{GUE5}.
This scaling is not applicable in quantum mechanics 
since $\left[\hat{r}_i,\,\hat{p}_j\right]=i\hbar\gamma^{1/3}\delta_{ij}\neq i\hbar\delta_{ij}$
holds. However, it is possible to define a scaled quantum Hamiltonian
by substituting $\hat{\boldsymbol{r}}=\gamma^{2/3}\boldsymbol{r}$
in the Schr\"odinger equation and introducing the scaled energy $\hat{\boldsymbol{E}}=\gamma^{-2/3}\boldsymbol{E}$.

We will now apply this scaling to the exciton system.
Let us write the Hamiltonian of excitons~(\ref{eq:H}) in the form
\begin{eqnarray}
H & = & -\frac{e^{2}}{4\pi\varepsilon_{0}\varepsilon}\frac{1}{r}- e\boldsymbol{F}\cdot\boldsymbol{r}
\nonumber \\
& + & H_0+(eB)H_1+(eB)^2 H_2 ,
\end{eqnarray}
with the $H_i$ given in Appendix~\ref{sub:Hamiltonianrm}. Due to the effective
masses of electron and hole and due to the scaling of 
the Coulomb energy by the dielectric constant,
we introduce exciton Hartree units so that 
the hydrogen-like part of the Hamiltonian
is exactly of the same form as that of the hydrogen 
Hamiltonian in normal Hartree units~\cite{50} (see Appendix~\ref{sub:Scaled-Hartree-units}).
Variables in exciton Hartree units will be indicated by a tilde sign.

Performing the substitution $\hat{\boldsymbol{r}}=\gamma^{2/3}\tilde{\boldsymbol{r}}/\alpha$
in the corresponding Schr\"odinger equation,
where we now have to use $\gamma=B/B_0$ with
$B_0=2.3505\times 10^5\,\mathrm{T}/\left(\gamma_1'^2\varepsilon^2\right)$, 
and multiplying the resulting equation with $\alpha^2\gamma^{2/3}$, we obtain
\begin{eqnarray}
& & -\frac{\alpha}{\hat{r}}- \gamma^{4/3}\alpha^3 \tilde{\boldsymbol{F}}\cdot\hat{\boldsymbol{r}}\nonumber \\
\nonumber \\
& + & \gamma^{2/3}\tilde{H}_0+\gamma^{1/3}\alpha^2 \tilde{H}_1+\alpha^4 \tilde{H}_2 \nonumber \\
\nonumber \\
& = & \gamma^{-2/3}\alpha^2 \tilde{E}.\label{eq:scaled}
\end{eqnarray}
As for the hydrogen atom, we define the scaled energy $\hat{E}=\gamma^{-2/3}\tilde{E}$
and scaled electric field strength $\hat{\boldsymbol{F}}=\gamma^{4/3}\tilde{\boldsymbol{F}}$.
When 
using the complete basis of Eq.~(\ref{eq:ansatz}),
Eq.~(\ref{eq:scaled}) represents a quadratic eigenvalue problem
of the form
\begin{equation}
\boldsymbol{A}\boldsymbol{c}+\tau\boldsymbol{B}\boldsymbol{c}=\tau^2 \boldsymbol{C}\boldsymbol{c}\label{eq:qgev}
\end{equation}
with hermitian matrices $\boldsymbol{A}$, $\boldsymbol{B}$, and $\boldsymbol{C}$
and an eigenvalue $\tau=\gamma^{1/3}$.
The eigenvalue problem can be changed to a standard generalized eigenvalue problem
by defining a vector $\boldsymbol{d}=\tau\boldsymbol{c}$:
\begin{equation}
\left(\begin{array}{cc}
\boldsymbol{A} & \boldsymbol{B}\\
\boldsymbol{0} & \boldsymbol{1}
\end{array}\right)\left(\begin{array}{c}
\boldsymbol{c}\\
\boldsymbol{d}
\end{array}\right)=\tau\left(\begin{array}{cc}
\boldsymbol{0} & \boldsymbol{C}\\
\boldsymbol{1} & \boldsymbol{0}
\end{array}\right)\left(\begin{array}{c}
\boldsymbol{c}\\
\boldsymbol{d}
\end{array}\right).
\end{equation}
This eigenvalue problem is solved for constant scaled energies $\hat{E}$
using an appropriate LAPACK routine~\cite{Lapack}.

We finally note that due to the 
substitution $\hat{\boldsymbol{r}}=\gamma^{2/3}\tilde{\boldsymbol{r}}/\alpha$
and due to the use of exciton Hartree units,
a different value of the free convergence parameter $\alpha$ than in Sec.~\ref{sec:Theory} has to be used
to obtain convergence for the exciton states with principal quantum number
$n$. This value is given by $\alpha\approx n\gamma^{2/3}$.

\section{Discussion of antiunitary symmetries\label{sec:analytical}}

In a previous paper~\cite{175} we have shown analytically
that the last remaining antiunitary symmetry known
from the hydrogen atom in external fields is broken
for the exciton Hamiltonian~(\ref{eq:H}) for most orientations of the external
fields. For the reader's convenience we recapitulate 
the most important steps as some of the results are important
for the following discussions.

The matrices $\boldsymbol{I}_{i}$ of the quasi-spin $I=1$
given by Eq.~(\ref{eq:I})
are not the standard spin matrices $\boldsymbol{S}_{i}$ 
of spin one~\cite{Messiah2}. 
However, these matrices
obey the commutation rules~\cite{25}
\begin{equation}
\left[\boldsymbol{I}_{i},\,\boldsymbol{I}_{j}\right]=
i\hbar\sum_{k=1}^{3}\varepsilon_{ijk}\boldsymbol{I}_{k},
\end{equation}
for which reason a unitary transformation 
can be found so that $\boldsymbol{U}^{\dagger}
\boldsymbol{I}_{i}\boldsymbol{U}=\boldsymbol{S}_{i}$ holds.
Since in Ref.~\cite{Messiah2} the behavior of the standard spin
matrices under symmetry operations such as
time reversal and reflections are given, 
we will use the matrices $\boldsymbol{S}_{i}$
instead of the $\boldsymbol{I}_{i}$ in the following.

In the special case of vanishing 
Luttinger parameters $\gamma_2=\gamma_3=0$, the exciton
Hamiltonian~(\ref{eq:H})
is of the same form as the Hamiltonian of a hydrogen atom
in external fields. 
It is well known that for this Hamiltonian
there is still one antiunitary symmetry left, i.e., that it
is invariant under the combined symmetry 
of time inversion $K$ followed by a reflection $S_{\hat{\boldsymbol{n}}}$ 
at the specific plane spanned by both fields~\cite{QSC}.
This plane is given by the normal vector 
\begin{equation}
\hat{\boldsymbol{n}}=\left(\boldsymbol{B}\times\boldsymbol{F}\right)/\left|\boldsymbol{B}\times\boldsymbol{F}\right|\label{eq:nvec}
\end{equation}
or $\hat{\boldsymbol{n}}\perp\hat{\boldsymbol{B}}=\boldsymbol{B}/B$ if $\boldsymbol{F}=\boldsymbol{0}$ holds.
Therefore, the hydrogen-like system shows GOE statistics in the chaotic regime.

As the hydrogen atom is spherically symmetric in the field-free case,
it makes no difference whether the magnetic field is
oriented in $z$ direction or not. 
However, in a semiconductor with $\delta'\neq 0$ the Hamiltonian
has cubic symmetry and the orientation
of the external fields with respect to the crystal axis
of the lattice becomes important.
Any rotation of the coordinate system with the aim 
of making the $z$ axis coincide with the direction of the magnetic
field will also rotate the cubic crystal lattice.
The only remaining antiunitary symmetry
mentioned above is now broken for the exciton Hamiltonian
if the plane spanned by both fields
is \emph{not} identical to one of the symmetry planes
of the cubic lattice.
Even without an external electric field the symmetry is broken
if the magnetic field is not oriented in 
one of these symmetry planes.
Only if the plane spanned by both fields
is identical to one of the symmetry planes of the cubic lattice,
the antiunitary symmetry $KS_{\hat{\boldsymbol{n}}}$
with $\hat{\boldsymbol{n}}$ given by Eq.~(\ref{eq:nvec})
is present since only then the reflection $S_{\hat{\boldsymbol{n}}}$
transforms the lattice into itself.

This criterion can also be expressed in a different way:
The antiunitary symmetry known from the hydrogen atom
is broken if none of the normal vectors $\hat{\boldsymbol{n}}_i$ of the $9$ symmetry
planes of the cubic lattice given by
\begin{eqnarray}
\hat{\boldsymbol{n}}_{1} & = & \left(1,\,0,\,0\right)^{\mathrm{T}},\nonumber \\
\hat{\boldsymbol{n}}_{2} & = & \left(0,\,1,\,0\right)^{\mathrm{T}},\nonumber \\
\hat{\boldsymbol{n}}_{3} & = & \left(0,\,0,\,1\right)^{\mathrm{T}},\nonumber \\
\hat{\boldsymbol{n}}_{4} & = & \left(1,\,1,\,0\right)^{\mathrm{T}}/\sqrt{2},\nonumber \\
\hat{\boldsymbol{n}}_{5} & = & \left(0,\,1,\,1\right)^{\mathrm{T}}/\sqrt{2},\nonumber \\
\hat{\boldsymbol{n}}_{6} & = & \left(1,\,0,\,1\right)^{\mathrm{T}}/\sqrt{2},\nonumber \\
\hat{\boldsymbol{n}}_{7} & = & \left(1,\,-1,\,0\right)^{\mathrm{T}}/\sqrt{2},\nonumber \\
\hat{\boldsymbol{n}}_{8} & = & \left(0,\,1,\,-1\right)^{\mathrm{T}}/\sqrt{2},\nonumber \\
\hat{\boldsymbol{n}}_{9} & = & \left(-1,\,0,\,1\right)^{\mathrm{T}}/\sqrt{2},\label{eq:ni}
\end{eqnarray}
is parallel to 
\begin{equation}
\hat{\boldsymbol{n}}=\left(\boldsymbol{B}\times\boldsymbol{F}\right)/\left|\boldsymbol{B}\times\boldsymbol{F}\right|,\label{eq:nvec}
\end{equation}
or, in the case of $\boldsymbol{F}=\boldsymbol{0}$,
if none of these vectors is perpendicular to
\begin{equation}
\hat{\boldsymbol{B}}=\boldsymbol{B}/B.\label{eq:Bvec}
\end{equation}

Since the breaking of all antiunitary symmetries depends on the 
relative orientation of the external fields to all normal vectors
$\hat{\boldsymbol{n}}_i$, we can introduce
a parameter which is a qualitative measure for the deviation from
the cases with antiunitary symmetry:
\begin{equation}
\sigma=\left[\sum_{i=1}^{9}\frac{\left|\boldsymbol{B}\times\boldsymbol{F}\right|^2}{\left|\hat{\boldsymbol{n}}_i\times\left(\boldsymbol{B}\times\boldsymbol{F}\right)\right|^2}\right]^{-\frac{1}{2}}.\label{eq:sigma1}
\end{equation}
For the special case of $\boldsymbol{F}=\boldsymbol{0}$ we define
\begin{equation}
\sigma=\left[\sum_{i=1}^{9}\left(\hat{\boldsymbol{n}}_i\cdot\hat{\boldsymbol{B}}\right)^{-2}\right]^{-\frac{1}{2}}.\label{eq:sigma}
\end{equation}
We have $\sigma=0$ for the cases with antiunitary symmetry; and that symmetry is more and more
broken with increasing values of $\sigma$.


Under time inversion $K$ and reflections $S_{\hat{\boldsymbol{n}}}$
at a plane perpendicular to a normal vector $\hat{\boldsymbol{n}}$
the vectors of position $\boldsymbol{r}$, momentum $\boldsymbol{p}$
and spin $\boldsymbol{S}$ transform according to~\cite{Messiah2}
\begin{subequations}
\begin{eqnarray}
K\boldsymbol{r}K^{\dagger}& = &\boldsymbol{r},\\
K\boldsymbol{p}K^{\dagger}& = &-\boldsymbol{p},\\
K\boldsymbol{S}K^{\dagger}& = &-\boldsymbol{S},
\end{eqnarray}
\end{subequations}
and
\begin{subequations}
\begin{eqnarray}
S_{\hat{\boldsymbol{n}}}\boldsymbol{r}S_{\hat{\boldsymbol{n}}}^{\dagger} & = & \boldsymbol{r}-2\hat{\boldsymbol{n}}\left(\hat{\boldsymbol{n}}\cdot\boldsymbol{r}\right),\\
S_{\hat{\boldsymbol{n}}}\boldsymbol{p}S_{\hat{\boldsymbol{n}}}^{\dagger} & = & \boldsymbol{p}-2\hat{\boldsymbol{n}}\left(\hat{\boldsymbol{n}}\cdot\boldsymbol{p}\right),\\
S_{\hat{\boldsymbol{n}}}\boldsymbol{S}S_{\hat{\boldsymbol{n}}}^{\dagger} & = & -\boldsymbol{S}+2\hat{\boldsymbol{n}}\left(\hat{\boldsymbol{n}}\cdot\boldsymbol{S}\right).
\end{eqnarray}
\end{subequations}

For all orientations of the external fields the hydrogen-like part of the Hamiltonian~(\ref{eq:H}) is
invariant under $KS_{\hat{\boldsymbol{n}}}$ with $\hat{\boldsymbol{n}}$ given by Eq.~(\ref{eq:nvec}). However, other parts
of the Hamiltonian such as $H_c=\left(p_{1}^{2}\boldsymbol{S}_{1}^{2}+\mathrm{c.p.}\right)$ [see Eq.~(\ref{eq:Hh})]
are not invariant if $\sigma\neq 0$ holds. 
For example, for the case with $\boldsymbol{B}\left(0,\,0\right)$ and $\boldsymbol{F}\left(\pi/6,\,\pi/2\right)$,
we obtain 
\begin{eqnarray}
& & S_{\hat{\boldsymbol{n}}}KH_{c}K^{\dagger}S_{\hat{\boldsymbol{n}}}^{\dagger}-H_{c}\nonumber\\
& = & 1/8\left[2\sqrt{3}\left(\boldsymbol{S}_{2}^{2}-\boldsymbol{S}_{1}^{2}\right)p_{1}p_{2}\right.\nonumber\\
& + & 3\left(\boldsymbol{S}_{1}^{2}p_{2}^{2}+\boldsymbol{S}_{2}^{2}p_{1}^{2}\right)-3\left(\boldsymbol{S}_{1}^{2}p_{1}^{2}+\boldsymbol{S}_{2}^{2}p_{2}^{2}\right)\nonumber\\
& + & \left.\left\{\boldsymbol{S}_{1},\boldsymbol{S}_{2}\right\}\left(2\sqrt{3}\left(p_{2}^{2}-p_{1}^{2}\right)+12p_{1}p_{2}\right)\right]\neq 0\label{eq:SKHCKS}
\end{eqnarray}
with $\hat{\boldsymbol{n}}=\left(-1/2,\,\sqrt{3}/2,\,0\right)^{\mathrm{T}}$.
Note that even though $H_c$ does not depend on the external fields,
the normal vector $\hat{\boldsymbol{n}}$ is determined 
by these fields via Eq.~(\ref{eq:nvec}). Otherwise, the
hydrogen-like part of the Hamiltonian would not be invariant under 
$KS_{\hat{\boldsymbol{n}}}$.

Since the expression in Eq.~(\ref{eq:SKHCKS})
is not equal to zero, we have shown for
$\boldsymbol{B}\left(0,\,0\right)$ and $\boldsymbol{F}\left(\pi/6,\,\pi/2\right)$
that the generalized time-reversal symmetry
of the hydrogen atom is broken
for excitons due to the cubic symmetry of the semiconductor.
The same calculation can also be performed for other orientations
of the external fields.
As we have stated above, the antiunitary symmetry remains
unbroken only for specific orientations of the fields.

\section{Appearance of GOE and GUE statistics\label{sub:appearance}}

We will now demonstrate the breaking of all antiunitary
symmetries by analyzing the nearest-neighbor spacings
of the energy eigenvalues corresponding to the 
Hamiltonian~(\ref{eq:H})~\cite{GUE1} for a model system with
the arbitrarily chosen set of parameters $E_{\mathrm{g}}=0$, $\varepsilon=7.5$, 
$m_{\mathrm{e}}=m_0$, $\gamma_1'=2$, $\mu'=0$, and $\delta'=-0.15$.
If we set $\boldsymbol{F}=\boldsymbol{0}$,
we expect to obtain GUE statistics in the limit of high energies 
as long as the magnetic field is not oriented in one of the symmetry planes
of the lattice.

Before analyzing the nearest-neighbor spacings,
we have to unfold the spectra to obtain 
a constant mean spacing~\cite{GUE1,QSC,QC_1,QC_16}.
The unfolding procedure separates the average behavior
of the non-universal spectral density from 
universal spectral fluctuations
and yields a spectrum in which the mean level spacing
is equal to unity~\cite{GUE4}.

\begin{figure*}
\begin{centering}
\includegraphics[width=2.0\columnwidth]{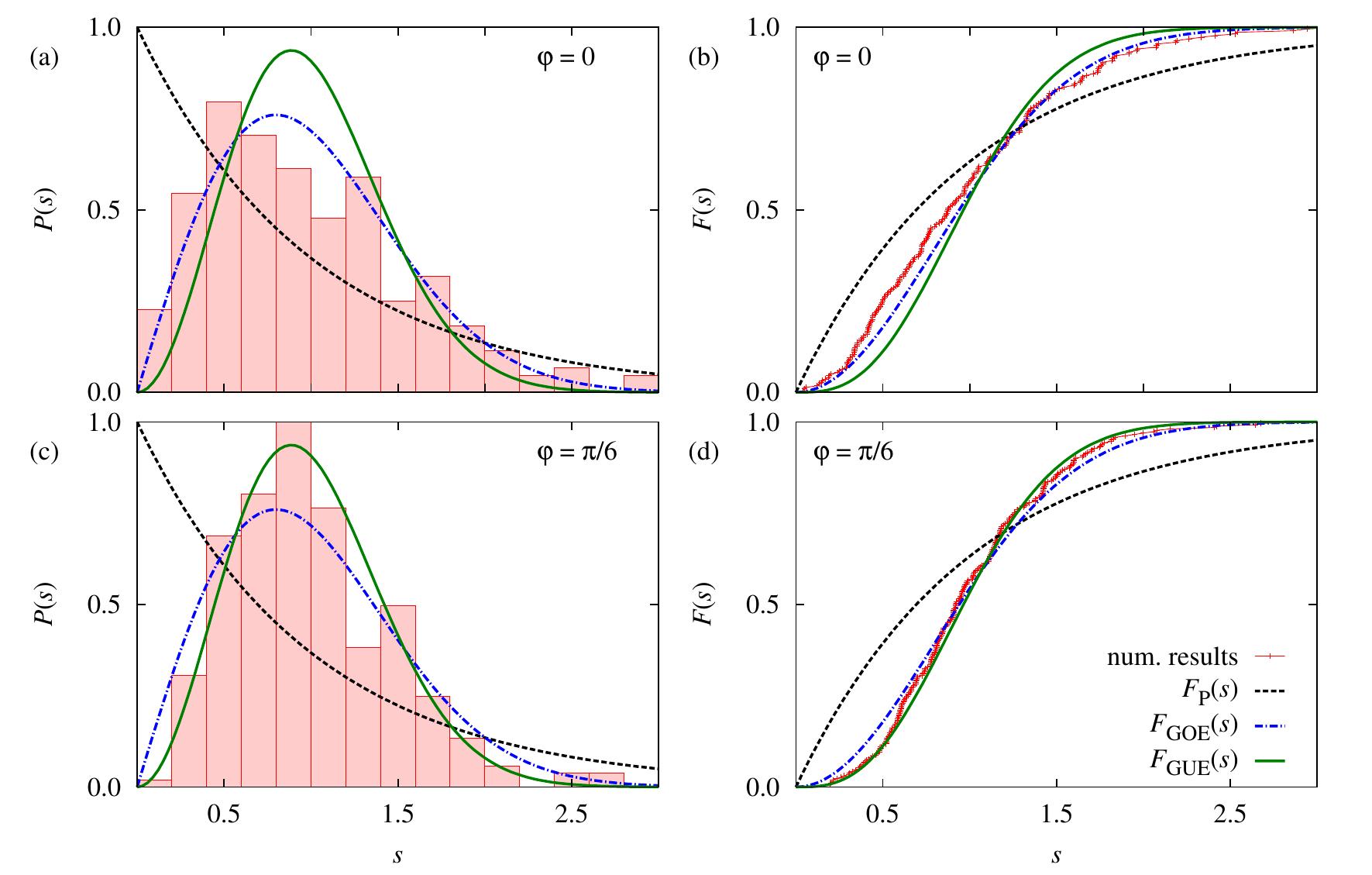}
\par\end{centering}

\protect\caption{Level
spacing probability distribution functions $P(s)$ (left)
and cumulative distribution functions $F\left(s\right)$ (right)
for $\delta'=-0.15$, $B=3\,\mathrm{T}$, $\vartheta=\pi/6$, 
and two different values of $\varphi$.
Besides the numerical data (red boxes or red dots), we also show the corresponding functions of a 
Poissonian ensemble (black dashed line), GOE (blue dash-dotted line),
and GUE (green solid line).
Only if the magnetic field is oriented in one of the
symmetry planes of the lattice, one antiunitary symmetry
is present and GOE statistics can be observed~(a,b).
In all other cases, all antiunitary symmetries are broken
and GUE statistics appears~(c,d).~\label{fig:numer}}
\end{figure*}

\begin{figure*}
\begin{centering}
\includegraphics[width=2.0\columnwidth]{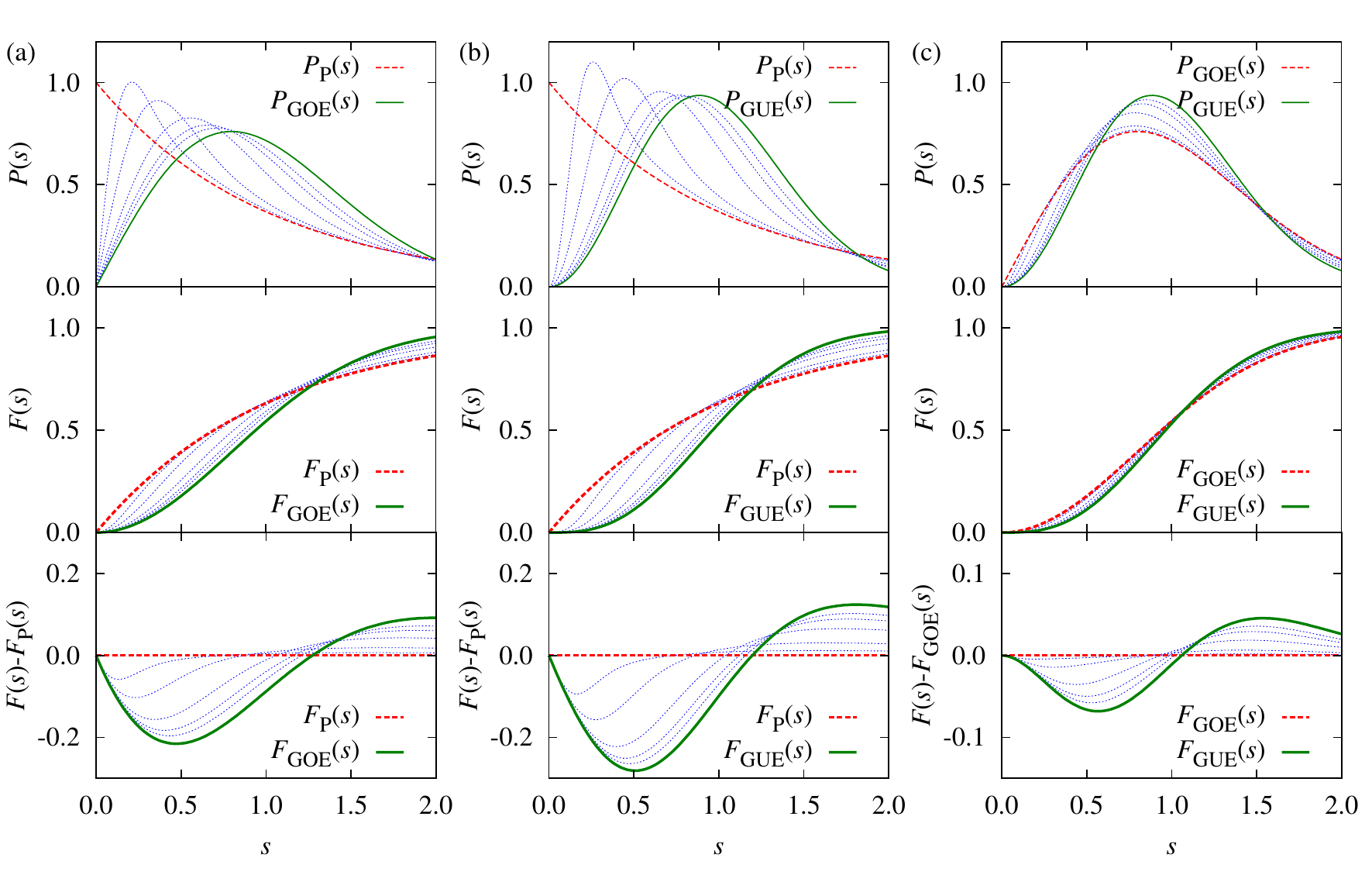}
\par\end{centering}

\protect\caption{(Color online) Level
spacing probability distribution functions $P(s)$ (first row)
and cumulative distribution functions $F\left(s\right)$ (second row)
for the transitions (a) $\mathrm{P}\rightarrow\mathrm{GOE}$, (b) $\mathrm{P}\rightarrow\mathrm{GUE}$,
and (c) $\mathrm{GOE}\rightarrow\mathrm{GUE}$. 
The blue dotted lines
show the transition functions of Eqs.~(\ref{eq:PGOE}), (\ref{eq:PGUE}), and~(\ref{eq:GOEGUE})
for $\lambda=0.1,\,0.2,\,0.4,\,0.6,\,0.8$.
To visualize the differences between the cumulative distribution functions
more clearly, especially for the $\mathrm{GOE}\rightarrow\mathrm{GUE}$ transition,
we also show in the third row the difference
between the distributions for a fixed value of $\lambda$ 
and the initial distribution, respectively.
\label{fig:trans}}

\end{figure*}

\begin{figure*}
\begin{centering}
\includegraphics[width=2.0\columnwidth]{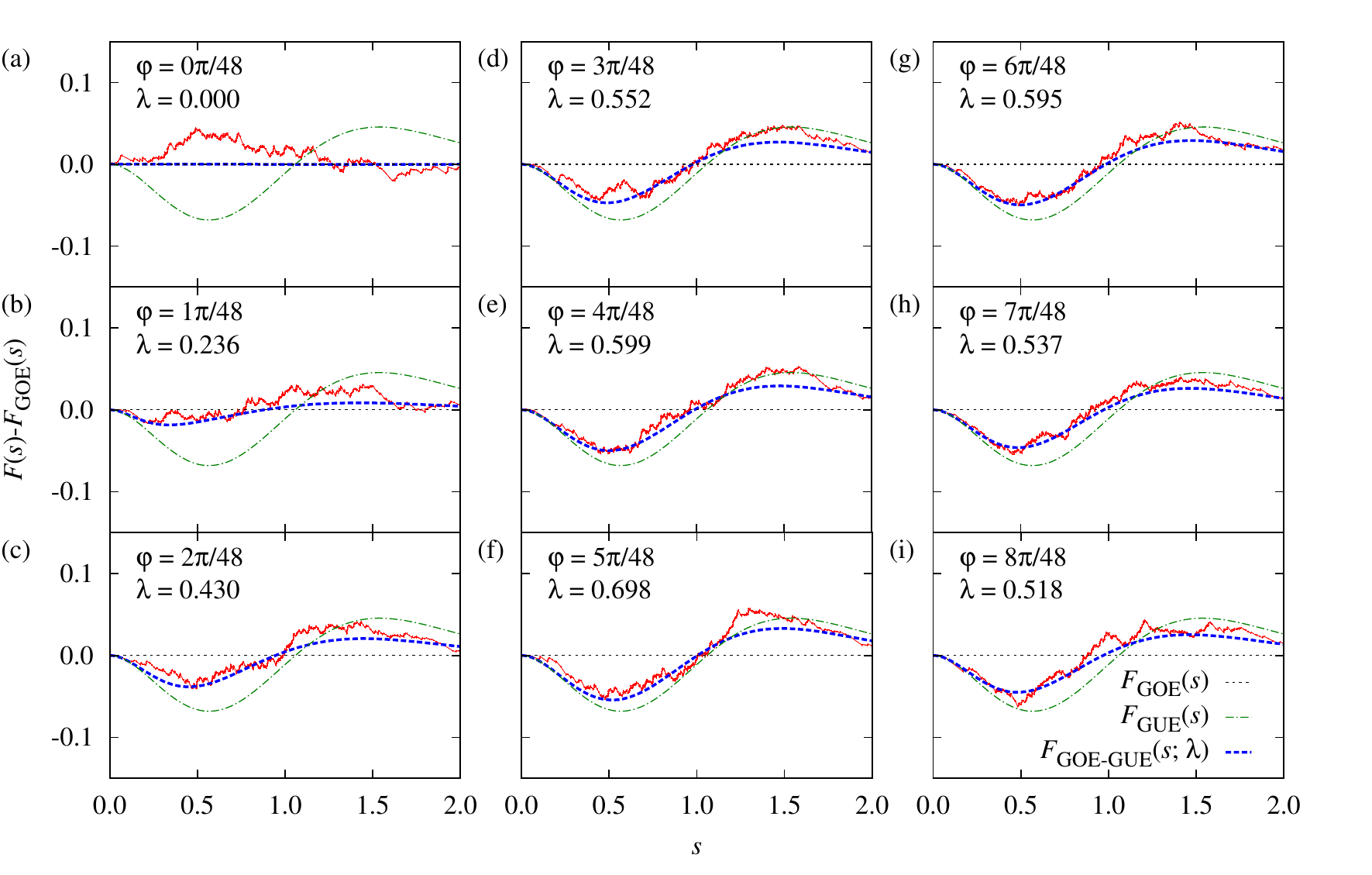}
\par\end{centering}

\protect\caption{Transition from GOE to GUE statistics
for fixed values of the magnetic field strength $B$ and increasing values
of the angle $\varphi$ in $\boldsymbol{B}\left(\varphi,\,\vartheta=\pi/6\right)$.
The results are presented in the same way as in the bottom most panel of Fig.~\ref{fig:trans}
to show the differences between $F_{\mathrm{GOE}}\left(s\right)$
and $F_{\mathrm{GUE}}\left(s\right)$ more clearly.
The data points (red) were fitted with the analytical
function $F_{\mathrm{GOE}\rightarrow\mathrm{GUE}}\left(s;\,\lambda\right)$.
The optimum values of the fit parameter $\lambda$ are given in each panel,
but also shown in Fig.~\ref{fig:phi2}.
One can observe a good agreement between the numerical data and the 
analytical function
describing the transition between the two statistics in dependence on $\lambda$.
Only for $\varphi=0$ the data shows a slight admixture of Poissonian statistics to the expected GOE statistics.
For further information see text.
\label{fig:phi}}

\end{figure*}




\begin{figure}
\begin{centering}
\includegraphics[width=1.0\columnwidth]{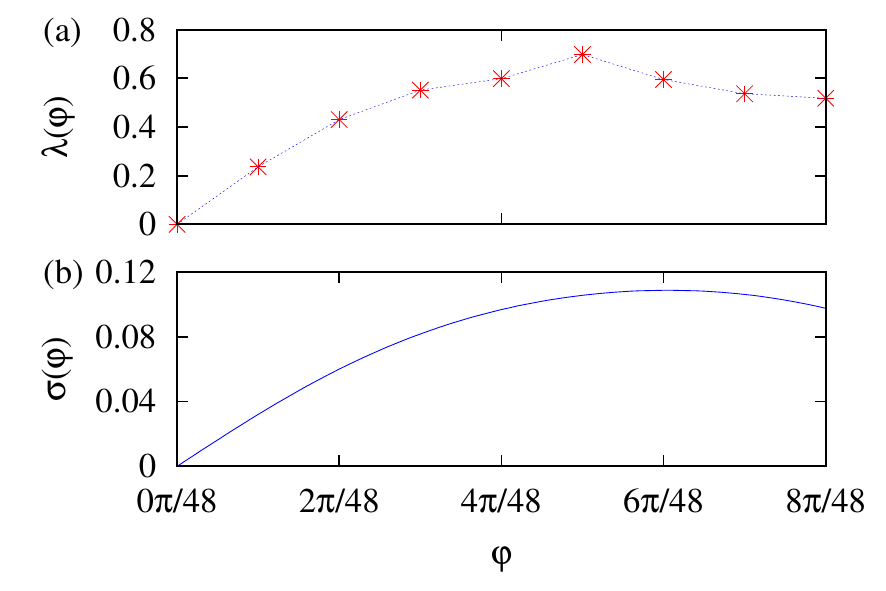}
\par\end{centering}

\protect\caption{(a) Optimum values of the fit parameter $\lambda$
in dependence on the angle $\varphi$ for the situation presented in Fig.~\ref{fig:phi}.
The blue dashed line only serves as a guide to the eye. 
(b) The function $\sigma\left(\varphi\right)$ of Eq.~(\ref{eq:sigma}) for $\vartheta=\pi/6$.
We obtain a qualitatively good agreement between both curves, i.e.,
as expected, both values $\lambda\left(\varphi\right)$ 
and $\sigma\left(\varphi\right)$ increase from zero to a certain value
and then decrease for $\varphi\gtrsim\pi/8$. 
\label{fig:phi2}}

\end{figure}

\begin{figure*}[t]
\begin{centering}
\includegraphics[width=2.0\columnwidth]{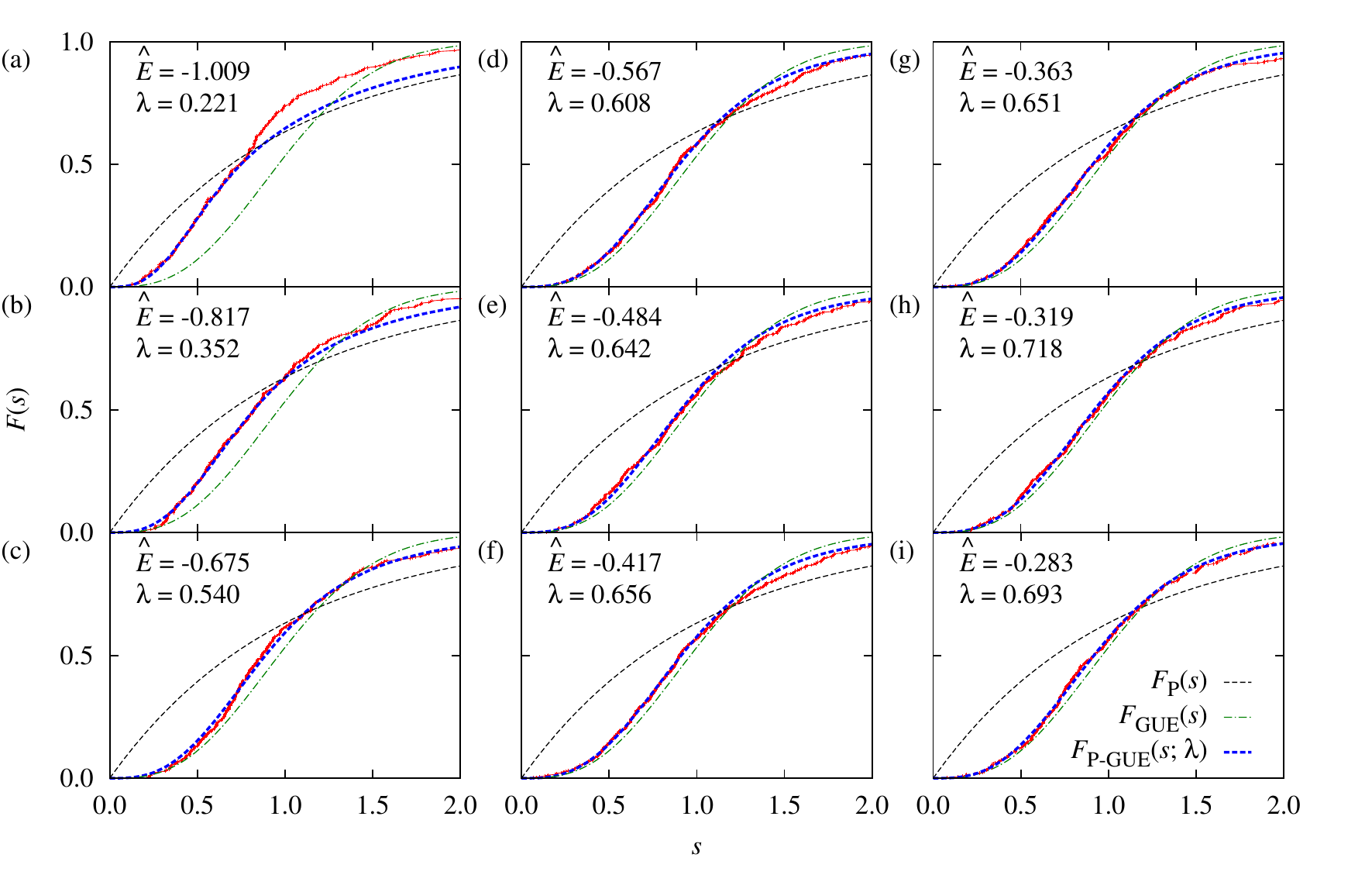}
\par\end{centering}

\protect\caption{Transition from Poissonian to GUE statistics
for fixed values of the angles $\varphi=\pi/8$, $\vartheta=\pi/6$ and increasing values
of the scaled energy $\hat{E}$.
Except for $\hat{E}=-1.009$ a good agreement between the numerical data and the analytical
function $F_{\mathrm{P}\rightarrow\mathrm{GUE}}\left(s;\,\lambda\right)$
is obtained. 
For further information see text.
\label{fig:Escal}}

\end{figure*}

\begin{figure}
\begin{centering}
\includegraphics[width=1.0\columnwidth]{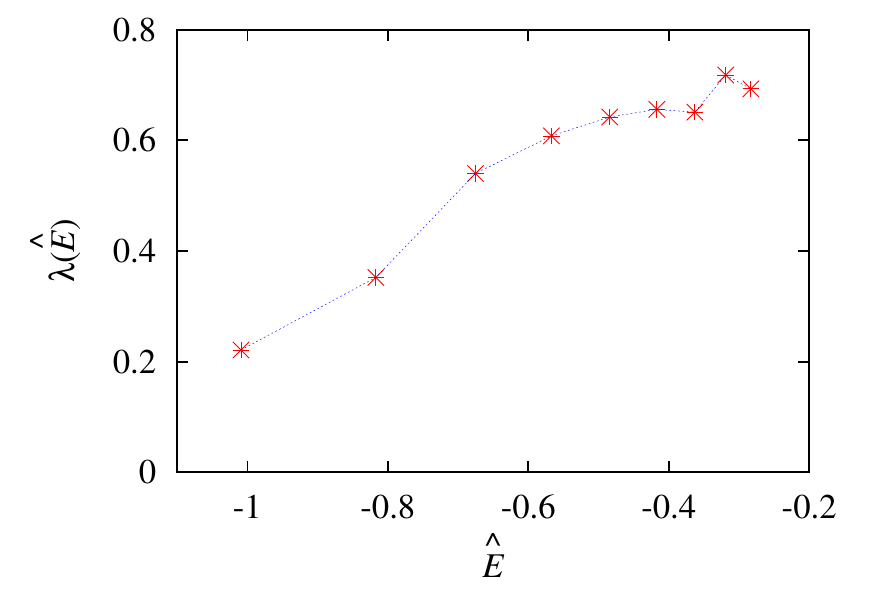}
\par\end{centering}

\protect\caption{Optimum values of the fit parameter $\lambda$
in dependence on the scaled energy $\hat{E}$ for the situation presented in Fig.~\ref{fig:Escal}.
The blue dashed line only serves as a guide to the eye. 
The value of $\lambda$ increases from a small value
at low scaled energies to about $\lambda\approx 0.7$, where 
the function $F_{\mathrm{P}\rightarrow\mathrm{GUE}}\left(s;\,\lambda\right)$
almost describes GUE statistics.
\label{fig:Escal2}}

\end{figure}

\textcolor{black}{To unfold the spectra, we plot
for the both cases of constant field 
strengths and of constant scaled energy
the number 
\begin{equation}
N(E)=\sum_{n}\Theta\left(E-E_n\right)
\end{equation}
of energy levels
up to the value $E_{\mathrm{max}}$, up 
to which all eigenvalues converged.
Here $\Theta(x)$ denotes the Heaviside function.
We leave out a certain number of low-lying sparse 
levels to remove individual but
nontypical fluctuations~\cite{GUE1}.
In the case of constant scaled energy
it is known that the mean number of levels 
is proportional to $E^{-2/3}$ in 
the dense part of the spectrum~\cite{GUE1}.
Hence, we fit $N(E)$ with $\bar{N}(E)=a E^{-2/3}+b$.
In the case of constant field strength no such 
proportionality is known and we fit $N(E)$ with 
a cubic polynomial function $\bar{N}(E)$.
The level spacings of the unfolded spectrum are then given by 
$s_n=\bar{N}(E_{n+1})-\bar{N}(E_{n})$~\cite{SPS}.}

Since the magnetic field
breaks all symmetries in the system and limits the
convergence of the solutions of the 
generalized eigenvalue problem with high energies~\cite{100},
the number of level spacings analyzed here
is comparatively small and comprises
about $250$ to $500$ exciton states.
In this case, the cumulative distribution function~\cite{GUE2}
\begin{equation}
F(s)=\int_{0}^{s}P(x)\,\mathrm{d}x
\end{equation}
is often more meaningful than
histograms of the level
spacing probability distribution function $P(s)$.

We will compare our results with the
distribution functions known from random
matrix theory~\cite{QC_1,QC}:
the Poissonian distribution 
\begin{equation}
P_{\mathrm{P}}(s)=e^{-s}
\end{equation}
for non-interacting energy levels, the Wigner distribution 
\begin{equation}
P_{\mathrm{GOE}}(s)=\frac{\pi}{2}\,se^{-\pi s^2/4},
\end{equation} 
and the distribution 
\begin{equation}
P_{\mathrm{GUE}}(s)=\frac{32}{\pi^2}\,s^2e^{-4 s^2/\pi}
\end{equation}
for systems without any antiunitary symmetry.
It can be seen that the most striking difference between the
three distributions is the behavior for small values of $s$.
While for the Poissonian distribution the probability of
level crossings in nonzero and thus $P_{\mathrm{P}}(0)\neq 0$ holds, 
in chaotic spectra the symmetry reduction leads to a correlation 
of levels and hence to a strong suppression of crossings.
Note that the most characteristic feature of GOE or GUE statistics is
the linear or quadratic level repulsion for small $s$, respectively.

In Fig.~\ref{fig:numer} we show the results for
level spacing probability distribution function
and the cumulative distribution function for
$\boldsymbol{B}\left(0,\,\pi/6\right)$ and
$\boldsymbol{B}\left(\pi/6,\,\pi/6\right)$
obtained with a constant magnetic field strength of $B=3\,\mathrm{T}$
and exciton states within a certain energy range.
While for $\boldsymbol{B}\left(0,\,\pi/6\right)$
the magnetic field is oriented in one of the symmetry planes of the 
lattice and thus only GOE statistics can be observed, we 
see clear evidence for GUE statistics as regards the
case with $\boldsymbol{B}\left(\pi/6,\,\pi/6\right)$.
Note that we have chosen the values $\delta'=-0.15$
and $B=3\,\mathrm{T}$ to be fixed.
It is well known from
atomic physics that chaotic effects become more apparent
in higher magnetic fields or by using states of higher energies for the
analysis. Hence, by increasing $B$ or investigating the statistics
of exciton states with higher energies, GUE statistics could
probably be observed also for smaller values of $|\delta'|$.
At this point we have to note that 
an evaluation of numerical spectra for $\delta'>0$ shows the same appearance of GUE statistics.
This is expected since the analytically shown breaking of all antiunitary symmetries in 
Sec.~\ref{sec:analytical} is independent of the sign of the material parameters.


\section{Transitions between spacing distributions\label{sub:transition}}

To the best of our knowledge, there are only two physical systems where both
the transition
from Poissonian to GUE statistics and the transition from GOE to GUE statistics
in dependence of a parameter of the system could be studied~\cite{QSC_K7_36,GUE4_21}.
As we have already stated in Secs.~\ref{sec:analytical} and~\ref{sub:appearance}, 
our system shows Poisson, GOE or GUE statistics in dependence on
the energy, the magnetic field strength and the angles 
$\vartheta$ and $\varphi$, i.e.,
in dependence of experimentally adjustable parameters.
Thus, our system is perfectly suited to investigate transitions between the
different statistics or different symmetry classes when changing one or more of these parameters.

In Ref.~\cite{GUE4} analytical expressions for the spacing distribution
functions in the transition region between the different statistics
have been derived using random matrix theory for $2\times 2$ matrices.
The transition from Poissonian to GOE statistics is described by
\begin{subequations}
\begin{eqnarray}
P_{\mathrm{P}\rightarrow\mathrm{GOE}}\left(s;\,\lambda\right) & = & Cse^{-D^{2}s^{2}}\nonumber \\
& \times & \int_{0}^{\infty}\,\mathrm{d}x\, e^{-x^{2}/4\lambda^{2}-x}I_{0}\left(z\right)
\end{eqnarray}
with $z=xDs/\lambda$ and
\begin{eqnarray}
D\left(\lambda\right) & = & \frac{\sqrt{\pi}}{2\lambda}U\left(-\frac{1}{2},\,0,\,\lambda^{2}\right),\\
\nonumber \\
C\left(\lambda\right) & = & 2D\left(\lambda\right)^{2},
\end{eqnarray}\label{eq:PGOE}%
\end{subequations}
a parameter $\lambda$, the Tricomi confluent hypergeometric function $U\left(a,\,b\,z\right)$~\cite{GUE4_35}
and the modified Bessel function $I_0\left(z\right)$~\cite{GUE4_35}.
For the special cases of $\lambda\rightarrow 0$ or $\lambda\rightarrow\infty$
Poissonian or GOE statistics is obtained, respectively.
However, already for $\lambda\gtrsim 0.7$ the transition to
GOE statistics is almost completed~\cite{GUE4}.

\textcolor{black}{At this point we have to note that 
the transition between different symmetry classes is not universal
and that the level spacing
distributions are universal only in the Poisson, GOE or GUE limit.
Besides the transition formula~(\ref{eq:PGOE}) derived within random matrix theory
also other interpolating distributions for the transition 
$\mathrm{P}\rightarrow\mathrm{GOE}$ have been proposed 
in the literature~\cite{GUE3_10a,GUE3_10b,GUE3_10c,GUE3_10d,GUE3_10e}.
When using one of these distributions for the intermediate
regime the results may be modified. 
However, since all the transition formulae presented here were derived in
the same manner within random matrix theory, we use these
formula for a consistent description of all transitions considered here.}

The transition from Poissonian to GUE statistics is described by
\begin{subequations}
\begin{eqnarray}
P_{\mathrm{P}\rightarrow\mathrm{GUE}}\left(s;\,\lambda\right) & = & Cs^{2}e^{-D^{2}s^{2}}\nonumber \\
& \times & \int_{0}^{\infty}\,\mathrm{d}x\, e^{-x^{2}/4\lambda^{2}-x}\frac{\sinh\left(z\right)}{z}
\end{eqnarray}
with $z=xDs/\lambda$ and
\begin{eqnarray}
D\left(\lambda\right) & = & \frac{1}{\sqrt{\pi}}+\frac{1}{2\lambda}e^{\lambda^{2}}\mathrm{erfc}\left(\lambda\right)-\frac{\lambda}{2}\mathrm{Ei}\left(\lambda^{2}\right)\nonumber \\
 & + & \frac{2\lambda^{2}}{\sqrt{\pi}}\,_{2}F_{2}\left(\frac{1}{2},\,1;\,\frac{3}{2},\,\frac{3}{2};\,\lambda^{2}\right),\\
\nonumber \\
C\left(\lambda\right) & = & \frac{4D\left(\lambda\right)^{3}}{\sqrt{\pi}},
\end{eqnarray}\label{eq:PGUE}%
\end{subequations}
the complementary error function erfc~\cite{GUE4_35}, the exponential integral Ei~\cite{GUE4_35}
and a generalized hypergeometric function $_2 F_2$~\cite{TI}.

Finally, the transition from GOE to GUE statistics is given by
\begin{subequations}
\begin{equation}
P_{\mathrm{GOE}\rightarrow\mathrm{GUE}}\left(s;\,\lambda\right) = Cse^{-D^{2}s^{2}}\mathrm{erf}\left(\frac{Ds}{\lambda}\right)
\end{equation}
with
\begin{eqnarray}
D\left(\lambda\right) & = & \frac{\sqrt{1+\lambda^{2}}}{\sqrt{\pi}}\left(\frac{\lambda}{1+\lambda^{2}}+\mathrm{arccot}\left(\lambda\right)\right),\\
\nonumber \\
C\left(\lambda\right) & = & 2\sqrt{1+\lambda^{2}}D\left(\lambda\right)^{2}.
\end{eqnarray}\label{eq:GOEGUE}%
\end{subequations}

As in Ref.~\cite{GUE4}, we calculate the distribution functions
for $\lambda=0.01\times 1000^{(k-1)/999}$ with $k=1,\ldots,1000$ and then
numerically integrate the results to obtain the corresponding 
cumulative distribution functions $F\left(s;\,\lambda\right)$.
All these functions are shown for different values of $\lambda$
in Fig.~\ref{fig:trans}.

As the transition from Poissonian to GOE statistics has been investigated
in detail for the hydrogen atom in external fields~\cite{GUE1},
we will treat the two other transitions in the following.

\subsection{GOE $\rightarrow$ GUE\label{sub:GOE-GUE}}

Let us start with the transition from GOE to GUE statistics.
For this case we solve the generalized eigenvalue problem~(\ref{eq:gev})
for different orientations of the magnetic field $\boldsymbol{B}\left(\varphi,\,\vartheta\right)$ by setting $\vartheta=\pi/6$ and gradually increasing the angle $\varphi$ from $0$ to $\pi/4$.
To increase the statistical significance, we analyze and merge the level spacings
for $B=2.8\,\mathrm{T}$, $B=3.0\,\mathrm{T}$, and $B=3.2\,\mathrm{T}$
for a given value of $\varphi$~\cite{GUE1}.
The results are finally fitted by the function
$F_{\mathrm{GOE}\rightarrow\mathrm{GUE}}\left(s;\,\lambda\right)$ and
shown in Fig.~\ref{fig:phi}.

For the special case of $\varphi=0$ we obtain GOE statistics
as expected since the magnetic field is oriented in the symmetry plane
of the solid with $\hat{\boldsymbol{n}}=\left(0,\,1,\,0\right)^{\mathrm{T}}$.
When increasing the angle $\varphi$, the parameter $\lambda$
changes rapidly from 0 to 0.5 and hence the transition from
GOE to GUE statistics is almost completed for $\varphi\gtrsim 3\pi/48$ (see Fig.~\ref{fig:phi2}).

The decrease of the parameter $\lambda$ for 
$\varphi\gtrsim\pi/8$ in Fig.~\ref{fig:phi2} can be explained by
considering the orientation of $\boldsymbol{B}$ with respect to all symmetry planes
of the lattice. Hence, we calculate the value of the parameter $\sigma$ of Eq.~(\ref{eq:sigma})
for $\vartheta=\pi/6$ and increasing values of $\varphi$.
It is obvious that the value of $\sigma$ increases for $0\leq\varphi\leq\pi/8$ and decreases
for $\pi/8\leq\varphi\leq\pi/4$ since the
magnetic field moves away from the plane with $\hat{\boldsymbol{n}}_{2}$
and then approaches the plane with $\hat{\boldsymbol{n}}_{7}$. Therefore, the fact that
$\boldsymbol{B}$ approaches the plane with $\hat{\boldsymbol{n}}_{7}$ for 
$\varphi\geq\pi/8$ explains the decrease of $\lambda$ in Fig.~\ref{fig:phi}.

\subsection{Poisson $\rightarrow$ GUE\label{sub:P-GUE}}

Let us now treat the transition from Poissonian to GUE statistics.
It is known from the hydrogen atom in external fields
that for fixed values of the magnetic field strength $B$
the low-energy part of the eigenvalue spectrum will show Poissonian statistics
while the high-energy part already shows GOE statistics.
For a better level statistics it is appropriate
to analyze the spectra with a constant scaled energy $\hat{E}$.

For fixed small values of the scaled energy the corresponding classical dynamics
becomes regular and energy eigenvalues of the quantum mechanical
system will show purely Poissonian statistics.
On the other hand, as we have shown above, GUE statistics is observed best at large energies
and for angles $\varphi$ and $\vartheta$, for which the magnetic field
is oriented exactly between two symmetry planes of the lattice.
Hence, keeping the values $\varphi=\pi/8$, $\vartheta=\pi/6$, and $\delta'=-0.15$ fixed
and increasing the scaled energy, we expect to observe a transition
from Poissonian to GUE statistics.

Having unfolded the spectra according to Ref.~\cite{GUE1},
we fit the numerical results by the function
$F_{\mathrm{P}\rightarrow\mathrm{GUE}}\left(s;\,\lambda\right)$ given in Eq.~(\ref{eq:PGUE}).
It can be seen from Fig.~\ref{fig:Escal} that
we obtain a good agreement
between the results for our system and the analytical function
for all scaled energies $\hat{E}>-0.9$.
The transition from Poissonian to GUE statistics
takes place already at very small values of the scaled energy $-1.2\lesssim\hat{E}\lesssim -0.6$
(see Fig.~\ref{fig:Escal2}).
This differs from the hydrogen atom in external fields where
the statistics is still Poisson-like for $\hat{E}\lesssim -0.6$~\cite{GUE1}
and can be explained by the presence of the cubic band structure here.
Therefore, the presence of the cubic band structure increases
the chaos in comparison with the hydrogen atom.

\textcolor{black}{
For very small values of the scaled energy $\hat{E}\lesssim -0.8$ a
reasonable analysis of the spectra is hardly possible.
For these values of $\hat{E}$ we cannot obtain enough converged 
eigenvalues in the dense part of the spectrum
due to the required computer memory.
On the other hand, the 
number of low-lying
sparse levels increases. Hence, fitting the number $N(E)$ of energy levels
with the function $\bar{N}(E)=a E^{-2/3}+b$ for the 
unfolding procedure (cf.~Sec.~\ref{sub:appearance})
does not lead to good results 
since the mean number of energy levels 
is proportional to $E^{-2/3}$
only in the dense part of the spectrum.
This effect can already be observed for $\hat{E}=-1.009$
in Fig.~\ref{fig:Escal}.
Note that a change in the unfolding procedure or the fit function 
would not lead to better results as the problem is connected with the
appearance of the low-lying
sparse levels. These levels lead to 
individual but nontypical fluctuations~\cite{GUE1}.}

It is generally assumed that the NNS
of large random matrices can be approximated by the NNS of
$2\times 2$ matrices of the same universality class~\cite{GUE4}.
Since we obtained a good agreement when fitting the 
functions~$F_{\mathrm{GOE}\rightarrow\mathrm{GUE}}\left(s;\,\lambda\right)$
and $F_{\mathrm{P}\rightarrow\mathrm{GUE}}\left(s;\,\lambda\right)$, 
which were derived for $2\times 2$ matrices,
to our numerical results,
we could prove the Wigner surmise~\cite{GUE4_8} for our system.

\section{Summary and outlook\label{sec:Summary}}

Investigating the Hamiltonian of excitons in cubic semiconductors
we could show analytically and numerically that the simultaneous presence
of the cubic band structure and external fields
can break all antiunitary symmetries in the system.
The level spacing statistics of the quantum mechanical spectrum
depends on the energy, the field strengths, the field orientations and on
the value of the parameter $\delta'$, which determines the strength of the
cubic deformation of the band structure.
This makes excitons in external fields a prime system to
investigate the transitions between different level spacing statistics.
Keeping the parameter $\delta'$ fixed, we analyzed the transition from 
GOE to GUE statistics and from Poissonian to GUE statistics.
A comparison with analytical formulae for these transitions derived
for $2\times 2$ matrices within random matrix theory
showed very good agreements. Hence, we could confirm the Wigner surmise
for our model system.

Since we changed only parameters such as the angles of the magnetic field
or the scaled energy, which can also be varied in experiments, we
think that the transition between the different level statistics 
could also be investigated experimentally. 
However, changing the two parameters $\delta'$ and the scaled energy $\hat{E}$
in numerical calculations will allow us to investigate arbitrary
transitions of the level statistics in the triangle between
Poissonian (arbitrary $\delta'$, small $\hat{E}$), 
GOE ($\delta'=0$, large $\hat{E}$), and GUE statistics 
($\delta'\neq 0$, large $\hat{E}$) in the future.
As for arbitrary transitions within this triangle
no analytical formulae have been derived within random matrix theory
so far, the corresponding functions $P\left(s;\,\lambda_1,\,\lambda_2\right)$
also have to be found.

\textcolor{black}{
We want to note that all transitions considered
here are modelled by Hamiltonians of the form 
$H=H_{\beta}+\lambda H_{\beta'}$~\cite{GUE4}, where 
$H_{\beta'}$ has a lower symmetry than $H_{\beta}$.
The level statistics is strongly affected by the perturbation $H_{\beta'}$
if the level spacings of $H_{\beta}$, which are 
smaller than the matrix elements of this 
Hamiltonian, and the matrix elements of $\lambda H_{\beta'}$
are of comparable size.
In the case of $\hbar\rightarrow 0$, the transition
will take place at even smaller values of $\lambda$.
Especially, the connection between $\lambda$ and the parameter $\sigma$ 
[cf.~Eqs.~(\ref{eq:sigma1}) and (\ref{eq:sigma})] must depend on $\hbar$.
However, we note that the parameter $\sigma$ has only been introduced
phenomenologically to describe the dependency of the transition on the angle
between the vector $\hat{\boldsymbol{n}}$~(\ref{eq:nvec}) 
or $\hat{\boldsymbol{B}}$~(\ref{eq:Bvec})
and the normal vectors $\hat{\boldsymbol{n}}_i$ 
of the symmetry planes of the lattice.}

\textcolor{black}{
To investigate the dependence of all results on $\hbar$, further and more extensive
calculations are necessary, which is beyond the scope of this work.
Nevertheless, our model system
offers the possibility for an according analysis and we will
discuss the effects in a future publication.}

Finally, we are certain that the discovery of 
GUE statistics for giant Rydberg 
excitons may pave the way to a deeper understanding of the 
connection between quantum and classical chaos.

\acknowledgments
F.S.~is  grateful  for  support  from  the
Landesgraduiertenf\"orderung of the Land Baden-W\"urttemberg.
We thank D.~Fr\"ohlich, M.~A{\ss}mann and M.~Bayer for helpful discussions.

\appendix

\begin{table*}

\protect\caption{Exciton Hartree units converted to SI-units 
for $\gamma_1'=2$ and $\varepsilon=7.5$. For a comparison, we also give the 
values for normal Hartree units, which are obtained by setting $\gamma_1'=\varepsilon=1$.\label{tab:1}}

\begin{centering}
\begin{tabular}{lllll}
\hline 
quantity\hspace{2cm} & symbol\hspace{0.5cm} & exc. Hartree unit\hspace{0.5cm} & SI $\left(\gamma_{1}'=2,\,\varepsilon=7.5\right)$\hspace{1cm} & SI $\left(\gamma_{1}'=1,\,\varepsilon=1\right)$\tabularnewline
\hline 
\hline 
charge & $q$ & $e$ & $1.6022\times10^{-19}$~C & $1.6022\times10^{-19}$~C\tabularnewline
action & $S$ & $\hbar$ & $1.0546\times10^{-34}$~Js & $1.0546\times10^{-34}$~Js\tabularnewline
mass & $m$ & $m_{0}/\gamma_{1}'$ & $4.5547\times10^{-31}$~kg & $9.1094\times10^{-31}$~kg\tabularnewline
length & $r$ & $\gamma_{1}'\varepsilon a_{0}$ & $7.9377\times10^{-10}$~m & $5.2918\times10^{-11}$~m\tabularnewline
momentum & $p$ & $\hbar/\gamma_{1}'\varepsilon a_{0}$ & $1.3286\times10^{-25}$~kg~m/s & $1.9929\times10^{-24}$~kg~m/s\tabularnewline
time & $t$ & $\gamma_{1}'\varepsilon^{2}a_{0}^{2}m_{0}/\hbar$ & $2.7213\times10^{-15}$~s & $2.4189\times10^{-17}$~s\tabularnewline
energy & $E$ & $\hbar^{2}/\gamma_{1}'\varepsilon^{2}a_{0}^{2}m_{0}$ & $3.8753\times10^{-20}$~J & $4.3597\times10^{-18}$~J\tabularnewline
magn. flux density & $B$ & $\hbar/\gamma_{1}'^{2}\varepsilon^{2}a_{0}^{2}e$ & $1.0447\times10^{+3}$~T & $2.3505\times10^{+5}$~T\tabularnewline
el. field strength & $F$ & $\hbar^{2}/\gamma_{1}'^{2}\varepsilon^{3}a_{0}^{3}m_{0}e$ & $3.0472\times10^{+8}$~V/m & $5.1422\times10^{+11}$~V/m\tabularnewline
\hline 
\end{tabular}
\par\end{centering}

\end{table*}

\section{Hamiltonian \label{sub:Hamiltonianrm}}

In this section we give the complete Hamiltonian of Eq.~(\ref{eq:H})
and describe the rotation necessary to make the quantization axis coincide 
with the direction of the magnetic field.
Let us write the Hamiltonian~(\ref{eq:H}) in the form
\begin{eqnarray}
H & = & E_{\mathrm{g}}-\frac{e^{2}}{4\pi\varepsilon_{0}\varepsilon}\frac{1}{r}\nonumber \\
\nonumber \\
& + & H_0+(eB)H_1+(eB)^2 H_2 - e\boldsymbol{F}\cdot\boldsymbol{r}
\end{eqnarray}
with $B=\left|\boldsymbol{B}\right|$. Using
$\hat{B}_i=B_i/B$ with the components $B_i$ of $\boldsymbol{B}$, the terms
$H_0$, $H_1$, and $H_2$ are given by

\begin{eqnarray}
H_{0} & = & \frac{1}{2m_{0}}\left(\gamma_{1}'+4\gamma_{2}\right)\boldsymbol{p}^{2}-\frac{3\gamma_{2}}{\hbar^{2}m_{0}}\left[\boldsymbol{I}_{1}^{2}p_{1}^{2}+\mathrm{c.p.}\right]\nonumber\\
\nonumber\\
& - & \frac{6\gamma_{3}}{\hbar^{2}m_{0}}\left[\left\{ \boldsymbol{I}_{1},\,\boldsymbol{I}_{2}\right\} p_{1}p_{2}+\mathrm{c.p.}\right],
\end{eqnarray}

\begin{eqnarray}
H_{1} & = & \frac{1}{2m_{0}}\left(\frac{2m_{0}}{m_{\mathrm{e}}}-\gamma_{1}'+4\gamma_{2}\right)\hat{\boldsymbol{B}}\cdot\boldsymbol{L}\nonumber\\
\nonumber\\
& + & \frac{3\gamma_{2}}{\hbar^{2}m_{0}}\left[\boldsymbol{I}_{1}^{2}\left(\hat{B}_{2}r_{3}p_{1}-\hat{B}_{3}r_{2}p_{1}\right)+\mathrm{c.p.}\right]\nonumber\\
\nonumber\\
& + & \frac{3\gamma_{3}}{\hbar^{2}m_{0}}\left[\left\{ \boldsymbol{I}_{1},\,\boldsymbol{I}_{2}\right\} \left(\hat{B}_{2}r_{3}p_{2}-\hat{B}_{1}r_{3}p_{1}\right.\right.\nonumber\\
\nonumber\\
 & & \qquad\quad \left.\left.+\hat{B}_{3}r_{1}p_{1}-\hat{B}_{3}r_{2}p_{2}\right)+\mathrm{c.p.}\right],
\end{eqnarray}

\begin{eqnarray}
H_{2} & = & \frac{1}{8m_{0}}\left(\gamma_{1}'+4\gamma_{2}\right)\left[\hat{\boldsymbol{B}}^{2}\boldsymbol{r}^{2}-\left(\hat{\boldsymbol{B}}\cdot\boldsymbol{r}\right)^{2}\right]\nonumber\\
\nonumber\\
& - & \frac{3\gamma_{2}}{4\hbar^{2}m_{0}}\left[\boldsymbol{I}_{1}^{2}\left(\hat{B}_{2}r_{3}-\hat{B}_{3}r_{2}\right)^{2}+\mathrm{c.p.}\right]\nonumber\\
\nonumber\\
& - & \frac{3\gamma_{3}}{2\hbar^{2}m_{0}}\left[\left\{ \boldsymbol{I}_{1},\,\boldsymbol{I}_{2}\right\} \left(\hat{B}_{2}r_{3}-\hat{B}_{3}r_{2}\right)\right.\nonumber\\
\nonumber\\
 & & \qquad\quad \left.\times\left(\hat{B}_{3}r_{1}-\hat{B}_{1}r_{3}\right)+\mathrm{c.p.}\right].
\end{eqnarray}

In our calculations, we express the magnetic field in spherical coordinates [see Eq.~(\ref{eq:spherical_coord})].
For the different orientations of the magnetic field 
we rotate the coordinate system by 
\begin{equation}
\boldsymbol{R}=\left(\begin{array}{ccc}
\cos\varphi\cos\vartheta & \sin\varphi\cos\vartheta & -\sin\vartheta\\
-\sin\varphi & \cos\varphi & 0\\
\cos\varphi\sin\vartheta & \sin\varphi\sin\vartheta & \cos\vartheta
\end{array}\right),
\end{equation}
i.e., we replace $\boldsymbol{x}\rightarrow\boldsymbol{x}'=\boldsymbol{R}^{\mathrm{T}}\boldsymbol{x}$ with $\boldsymbol{x}\in\left\{\boldsymbol{r},\,\boldsymbol{p},\,\boldsymbol{L},\,\boldsymbol{I},\,\boldsymbol{S}\right\}$
to make the quantization axis coincide with the direction of the magnetic
field~\cite{44,ED}.
Finally we express the Hamiltonian in terms of irreducible tensors (see, e.g., Refs.~\cite{ED,7_11,100,125})
and calculate the matrix elements of the matrices $\boldsymbol{D}$ and $\boldsymbol{M}$
in the generalized eigenvalue problem~(\ref{eq:gev}) or the matrices $\boldsymbol{A}$, $\boldsymbol{B}$,
and $\boldsymbol{C}$ in the generalized eigenvalue problem~(\ref{eq:qgev}).

\section{Exciton Hartree units \label{sub:Scaled-Hartree-units}}

When performing numerical calculations for the
hydrogen atom in external fields, often Hartree units are used~\cite{50,50_50}.
These units are obtained by setting the fundamental physical constants
$e$, $m_0$, $\hbar$ as well as the Bohr radius $a_{\mathrm{0}}$ to one.
As the effective masses of the electron and hole
differ from the free electron mass and since the Coulomb interaction
is scaled by the dielectric constant $\varepsilon$, we introduce
exciton Hartree units. Within these units the hydrogen-like part of the 
Hamiltonian~(\ref{eq:H}) is exactly of the same form as the
Hamiltonian of the hydrogen atom in Hartree units~\cite{50}
and the values of the scaled energies in Sec.~\ref{sec:scaled-energy}
can be compared directly with the values of the scaled energies
used in calculations for the hydrogen atom~\cite{GUE1}.
The exciton Hartree units are obtained by setting
$e=\hbar=1$, $m_0=\gamma_1'$ and $a_{\mathrm{exc}}=\gamma_1'\varepsilon a_0=1$.
Since all other physical quantities have to be converted to exciton
Hartree units as well, we give the according scaling factors
in Table~\ref{tab:1}.
Variables given in exciton Hartree units are marked by a tilde sign, e.g.,
$r\rightarrow\tilde{r}$, throughout the paper.


\begin{thebibliography}{82}
\expandafter\ifx\csname natexlab\endcsname\relax\def\natexlab#1{#1}\fi
\expandafter\ifx\csname bibnamefont\endcsname\relax
  \def\bibnamefont#1{#1}\fi
\expandafter\ifx\csname bibfnamefont\endcsname\relax
  \def\bibfnamefont#1{#1}\fi
\expandafter\ifx\csname citenamefont\endcsname\relax
  \def\citenamefont#1{#1}\fi
\expandafter\ifx\csname url\endcsname\relax
  \def\url#1{\texttt{#1}}\fi
\expandafter\ifx\csname urlprefix\endcsname\relax\def\urlprefix{URL }\fi
\providecommand{\bibinfo}[2]{#2}
\providecommand{\eprint}[2][]{\url{#2}}

\bibitem[{\citenamefont{Bohigas et~al.}(1984)\citenamefont{Bohigas, Giannoni,
  and Schmit}}]{QC_1}
\bibinfo{author}{\bibfnamefont{O.}~\bibnamefont{Bohigas}},
  \bibinfo{author}{\bibfnamefont{M.~J.} \bibnamefont{Giannoni}},
  \bibnamefont{and} \bibinfo{author}{\bibfnamefont{C.}~\bibnamefont{Schmit}},
  \bibinfo{journal}{Phys. Rev. Lett.} \textbf{\bibinfo{volume}{52}},
  \bibinfo{pages}{1} (\bibinfo{year}{1984}).

\bibitem[{\citenamefont{Mehta}(2004)}]{QSC_29}
\bibinfo{author}{\bibfnamefont{M.~L.} \bibnamefont{Mehta}},
  \emph{\bibinfo{title}{Random Matrices}} (\bibinfo{publisher}{Elsevier},
  \bibinfo{address}{Amsterdam}, \bibinfo{year}{2004}), \bibinfo{edition}{3rd}
  ed.

\bibitem[{\citenamefont{Porter}(1965)}]{QSC_30}
\bibinfo{editor}{\bibfnamefont{C.~E.} \bibnamefont{Porter}}, ed.,
  \emph{\bibinfo{title}{Statistical Theory of Spectra}}
  (\bibinfo{publisher}{Academic Press}, \bibinfo{address}{New York},
  \bibinfo{year}{1965}).

\bibitem[{\citenamefont{Rao and Taylor}(2002)}]{GUE5}
\bibinfo{author}{\bibfnamefont{J.}~\bibnamefont{Rao}} \bibnamefont{and}
  \bibinfo{author}{\bibfnamefont{K.~T.} \bibnamefont{Taylor}},
  \bibinfo{journal}{J. Phys. B: At. Mol. Opt. Phys.}
  \textbf{\bibinfo{volume}{35}}, \bibinfo{pages}{2627} (\bibinfo{year}{2002}).

\bibitem[{\citenamefont{Schierenberg et~al.}(2012)\citenamefont{Schierenberg,
  Bruckmann, and Wettig}}]{GUE4}
\bibinfo{author}{\bibfnamefont{S.}~\bibnamefont{Schierenberg}},
  \bibinfo{author}{\bibfnamefont{F.}~\bibnamefont{Bruckmann}},
  \bibnamefont{and} \bibinfo{author}{\bibfnamefont{T.}~\bibnamefont{Wettig}},
  \bibinfo{journal}{Phys. Rev. E} \textbf{\bibinfo{volume}{85}},
  \bibinfo{pages}{061130} (\bibinfo{year}{2012}).

\bibitem[{\citenamefont{Lenz and Haake}(1991)}]{GUE3}
\bibinfo{author}{\bibfnamefont{G.}~\bibnamefont{Lenz}} \bibnamefont{and}
  \bibinfo{author}{\bibfnamefont{F.}~\bibnamefont{Haake}},
  \bibinfo{journal}{Phys. Rev. Lett.} \textbf{\bibinfo{volume}{67}},
  \bibinfo{pages}{1} (\bibinfo{year}{1991}).

\bibitem[{\citenamefont{Haake}(2010)}]{QSC}
\bibinfo{author}{\bibfnamefont{F.}~\bibnamefont{Haake}},
  \emph{\bibinfo{title}{Quantum Signatures of Chaos}}, Springer Series in
  Synergetics (\bibinfo{publisher}{Springer}, \bibinfo{address}{Heidelberg},
  \bibinfo{year}{2010}), \bibinfo{edition}{3rd} ed.

\bibitem[{\citenamefont{Pandey}(1979)}]{GUE3_3}
\bibinfo{author}{\bibfnamefont{A.}~\bibnamefont{Pandey}},
  \bibinfo{journal}{Ann. Phys.} \textbf{\bibinfo{volume}{119}},
  \bibinfo{pages}{170} (\bibinfo{year}{1979}).

\bibitem[{\citenamefont{Mitchell et~al.}(2010)\citenamefont{Mitchell, Richter,
  and Weidenm\"{u}ller}}]{QC_5}
\bibinfo{author}{\bibfnamefont{G.~E.} \bibnamefont{Mitchell}},
  \bibinfo{author}{\bibfnamefont{A.}~\bibnamefont{Richter}}, \bibnamefont{and}
  \bibinfo{author}{\bibfnamefont{H.~A.} \bibnamefont{Weidenm\"{u}ller}},
  \bibinfo{journal}{Rev. Mod. Phys.} \textbf{\bibinfo{volume}{82}},
  \bibinfo{pages}{2845} (\bibinfo{year}{2010}).

\bibitem[{\citenamefont{Brody et~al.}(1981{\natexlab{a}})\citenamefont{Brody,
  Flores, French, Mello, Pandey, and Wong}}]{QSC_11}
\bibinfo{author}{\bibfnamefont{T.~A.} \bibnamefont{Brody}},
  \bibinfo{author}{\bibfnamefont{J.}~\bibnamefont{Flores}},
  \bibinfo{author}{\bibfnamefont{J.~B.} \bibnamefont{French}},
  \bibinfo{author}{\bibfnamefont{P.~A.} \bibnamefont{Mello}},
  \bibinfo{author}{\bibfnamefont{A.}~\bibnamefont{Pandey}}, \bibnamefont{and}
  \bibinfo{author}{\bibfnamefont{S.~S.~M.} \bibnamefont{Wong}},
  \bibinfo{journal}{Rev. Mod. Phys.} \textbf{\bibinfo{volume}{53}},
  \bibinfo{pages}{385} (\bibinfo{year}{1981}{\natexlab{a}}).

\bibitem[{\citenamefont{Rosenzweig and Porter}(1960)}]{QSC_12}
\bibinfo{author}{\bibfnamefont{N.}~\bibnamefont{Rosenzweig}} \bibnamefont{and}
  \bibinfo{author}{\bibfnamefont{C.~E.} \bibnamefont{Porter}},
  \bibinfo{journal}{Phys. Rev.} \textbf{\bibinfo{volume}{120}},
  \bibinfo{pages}{1698} (\bibinfo{year}{1960}).

\bibitem[{\citenamefont{Camarda and Georgopulos}(1983)}]{QSC_13}
\bibinfo{author}{\bibfnamefont{H.~S.} \bibnamefont{Camarda}} \bibnamefont{and}
  \bibinfo{author}{\bibfnamefont{P.~D.} \bibnamefont{Georgopulos}},
  \bibinfo{journal}{Phys. Rev. Lett.} \textbf{\bibinfo{volume}{50}},
  \bibinfo{pages}{492} (\bibinfo{year}{1983}).

\bibitem[{\citenamefont{St\"{o}ckmann and Stein}(1990)}]{QSC_15}
\bibinfo{author}{\bibfnamefont{H.-J.} \bibnamefont{St\"{o}ckmann}}
  \bibnamefont{and} \bibinfo{author}{\bibfnamefont{J.}~\bibnamefont{Stein}},
  \bibinfo{journal}{Phys. Rev. Lett.} \textbf{\bibinfo{volume}{64}},
  \bibinfo{pages}{2215} (\bibinfo{year}{1990}).

\bibitem[{\citenamefont{Alt et~al.}(1995)\citenamefont{Alt, Gr\"{a}f, Harney,
  Hofferbert, Lengeler, Richter, Schardt, and Weidenm\"{u}ller}}]{QSC_16}
\bibinfo{author}{\bibfnamefont{H.}~\bibnamefont{Alt}},
  \bibinfo{author}{\bibfnamefont{H.-D.} \bibnamefont{Gr\"{a}f}},
  \bibinfo{author}{\bibfnamefont{H.~L.} \bibnamefont{Harney}},
  \bibinfo{author}{\bibfnamefont{R.}~\bibnamefont{Hofferbert}},
  \bibinfo{author}{\bibfnamefont{H.}~\bibnamefont{Lengeler}},
  \bibinfo{author}{\bibfnamefont{A.}~\bibnamefont{Richter}},
  \bibinfo{author}{\bibfnamefont{P.}~\bibnamefont{Schardt}}, \bibnamefont{and}
  \bibinfo{author}{\bibfnamefont{H.~A.} \bibnamefont{Weidenm\"{u}ller}},
  \bibinfo{journal}{Phys. Rev. Lett.} \textbf{\bibinfo{volume}{74}},
  \bibinfo{pages}{62} (\bibinfo{year}{1995}).

\bibitem[{\citenamefont{Alt et~al.}(1996)\citenamefont{Alt, Gr\"{a}f,
  Hofferbert, Rangacharyulu, Rehfeld, Richter, Schardt, and Wirzba}}]{QSC_17}
\bibinfo{author}{\bibfnamefont{H.}~\bibnamefont{Alt}},
  \bibinfo{author}{\bibfnamefont{H.-D.} \bibnamefont{Gr\"{a}f}},
  \bibinfo{author}{\bibfnamefont{R.}~\bibnamefont{Hofferbert}},
  \bibinfo{author}{\bibfnamefont{C.}~\bibnamefont{Rangacharyulu}},
  \bibinfo{author}{\bibfnamefont{H.}~\bibnamefont{Rehfeld}},
  \bibinfo{author}{\bibfnamefont{A.}~\bibnamefont{Richter}},
  \bibinfo{author}{\bibfnamefont{P.}~\bibnamefont{Schardt}}, \bibnamefont{and}
  \bibinfo{author}{\bibfnamefont{A.}~\bibnamefont{Wirzba}},
  \bibinfo{journal}{Phys. Rev. E} \textbf{\bibinfo{volume}{54}},
  \bibinfo{pages}{2303} (\bibinfo{year}{1996}).

\bibitem[{\citenamefont{Zimmermann et~al.}(1988)\citenamefont{Zimmermann,
  K\"{o}ppel, Cederbaum, Persch, and Demtr\"{o}der}}]{QSC_18}
\bibinfo{author}{\bibfnamefont{T.}~\bibnamefont{Zimmermann}},
  \bibinfo{author}{\bibfnamefont{H.}~\bibnamefont{K\"{o}ppel}},
  \bibinfo{author}{\bibfnamefont{L.~S.} \bibnamefont{Cederbaum}},
  \bibinfo{author}{\bibfnamefont{G.}~\bibnamefont{Persch}}, \bibnamefont{and}
  \bibinfo{author}{\bibfnamefont{W.}~\bibnamefont{Demtr\"{o}der}},
  \bibinfo{journal}{Phys. Rev. Lett.} \textbf{\bibinfo{volume}{61}},
  \bibinfo{pages}{3} (\bibinfo{year}{1988}).

\bibitem[{\citenamefont{Zhou et~al.}(2010)\citenamefont{Zhou, Chen, Zhang, Yu,
  Lu, and Shen}}]{QC_3}
\bibinfo{author}{\bibfnamefont{W.}~\bibnamefont{Zhou}},
  \bibinfo{author}{\bibfnamefont{Z.}~\bibnamefont{Chen}},
  \bibinfo{author}{\bibfnamefont{B.}~\bibnamefont{Zhang}},
  \bibinfo{author}{\bibfnamefont{C.~H.} \bibnamefont{Yu}},
  \bibinfo{author}{\bibfnamefont{W.}~\bibnamefont{Lu}}, \bibnamefont{and}
  \bibinfo{author}{\bibfnamefont{S.~C.} \bibnamefont{Shen}},
  \bibinfo{journal}{Phys. Rev. Lett.} \textbf{\bibinfo{volume}{105}},
  \bibinfo{pages}{024101} (\bibinfo{year}{2010}).

\bibitem[{\citenamefont{Vina et~al.}(1998)\citenamefont{Vina, Potemski, and
  Wang}}]{QC_4}
\bibinfo{author}{\bibfnamefont{L.}~\bibnamefont{Vina}},
  \bibinfo{author}{\bibfnamefont{M.}~\bibnamefont{Potemski}}, \bibnamefont{and}
  \bibinfo{author}{\bibfnamefont{W.}~\bibnamefont{Wang}},
  \bibinfo{journal}{Phys.-Usp.} \textbf{\bibinfo{volume}{41}},
  \bibinfo{pages}{153} (\bibinfo{year}{1998}).

\bibitem[{\citenamefont{Held et~al.}(1998)\citenamefont{Held, Schlichter,
  Raithel, and Walther}}]{QSC_19}
\bibinfo{author}{\bibfnamefont{H.}~\bibnamefont{Held}},
  \bibinfo{author}{\bibfnamefont{J.}~\bibnamefont{Schlichter}},
  \bibinfo{author}{\bibfnamefont{G.}~\bibnamefont{Raithel}}, \bibnamefont{and}
  \bibinfo{author}{\bibfnamefont{H.}~\bibnamefont{Walther}},
  \bibinfo{journal}{Europhys. Lett.} \textbf{\bibinfo{volume}{43}},
  \bibinfo{pages}{392} (\bibinfo{year}{1998}).

\bibitem[{\citenamefont{Frisch et~al.}(2014)\citenamefont{Frisch, Mark, Aikawa,
  Ferlaino, Bohn, Makrides, Petrov, and Kotochigova}}]{QC_2}
\bibinfo{author}{\bibfnamefont{A.}~\bibnamefont{Frisch}},
  \bibinfo{author}{\bibfnamefont{M.}~\bibnamefont{Mark}},
  \bibinfo{author}{\bibfnamefont{K.}~\bibnamefont{Aikawa}},
  \bibinfo{author}{\bibfnamefont{F.}~\bibnamefont{Ferlaino}},
  \bibinfo{author}{\bibfnamefont{J.~L.} \bibnamefont{Bohn}},
  \bibinfo{author}{\bibfnamefont{C.}~\bibnamefont{Makrides}},
  \bibinfo{author}{\bibfnamefont{A.}~\bibnamefont{Petrov}}, \bibnamefont{and}
  \bibinfo{author}{\bibfnamefont{S.}~\bibnamefont{Kotochigova}},
  \bibinfo{journal}{Nature} \textbf{\bibinfo{volume}{507}},
  \bibinfo{pages}{475} (\bibinfo{year}{2014}).

\bibitem[{\citenamefont{Wintgen and Friedrich}(1987)}]{GUE1}
\bibinfo{author}{\bibfnamefont{D.}~\bibnamefont{Wintgen}} \bibnamefont{and}
  \bibinfo{author}{\bibfnamefont{H.}~\bibnamefont{Friedrich}},
  \bibinfo{journal}{Phys. Rev. A} \textbf{\bibinfo{volume}{35}},
  \bibinfo{pages}{1464(R)} (\bibinfo{year}{1987}).

\bibitem[{\citenamefont{Friedrich and Wintgen}(1989)}]{GUE9}
\bibinfo{author}{\bibfnamefont{H.}~\bibnamefont{Friedrich}} \bibnamefont{and}
  \bibinfo{author}{\bibfnamefont{D.}~\bibnamefont{Wintgen}},
  \bibinfo{journal}{Phys. Rep.} \textbf{\bibinfo{volume}{183}},
  \bibinfo{pages}{37} (\bibinfo{year}{1989}).

\bibitem[{\citenamefont{Berry and Tabor}(1977)}]{GUE3_6}
\bibinfo{author}{\bibfnamefont{M.~V.} \bibnamefont{Berry}} \bibnamefont{and}
  \bibinfo{author}{\bibfnamefont{M.}~\bibnamefont{Tabor}},
  \bibinfo{journal}{Proc. Roy. Soc. London A} \textbf{\bibinfo{volume}{356}},
  \bibinfo{pages}{375} (\bibinfo{year}{1977}).

\bibitem[{\citenamefont{So et~al.}(1995)\citenamefont{So, Anlage, Ott, and
  Oerter}}]{QSC_27}
\bibinfo{author}{\bibfnamefont{P.}~\bibnamefont{So}},
  \bibinfo{author}{\bibfnamefont{S.~M.} \bibnamefont{Anlage}},
  \bibinfo{author}{\bibfnamefont{E.}~\bibnamefont{Ott}}, \bibnamefont{and}
  \bibinfo{author}{\bibfnamefont{R.~N.} \bibnamefont{Oerter}},
  \bibinfo{journal}{Phys. Rev. Lett.} \textbf{\bibinfo{volume}{74}},
  \bibinfo{pages}{2662} (\bibinfo{year}{1995}).

\bibitem[{\citenamefont{Sacha et~al.}(1999)\citenamefont{Sacha, Zakrzewski, and
  Delande}}]{GUE8}
\bibinfo{author}{\bibfnamefont{K.}~\bibnamefont{Sacha}},
  \bibinfo{author}{\bibfnamefont{J.}~\bibnamefont{Zakrzewski}},
  \bibnamefont{and} \bibinfo{author}{\bibfnamefont{D.}~\bibnamefont{Delande}},
  \bibinfo{journal}{Phys. Rev. Lett.} \textbf{\bibinfo{volume}{83}},
  \bibinfo{pages}{2922} (\bibinfo{year}{1999}).

\bibitem[{\citenamefont{Shukla and Pandey}(1997)}]{GUE4_16}
\bibinfo{author}{\bibfnamefont{P.}~\bibnamefont{Shukla}} \bibnamefont{and}
  \bibinfo{author}{\bibfnamefont{A.}~\bibnamefont{Pandey}},
  \bibinfo{journal}{Nonlinearity} \textbf{\bibinfo{volume}{10}},
  \bibinfo{pages}{979} (\bibinfo{year}{1997}).

\bibitem[{\citenamefont{Haake et~al.}(1987)\citenamefont{Haake, Ku\'{s}, and
  Scharf}}]{QSC_K7_36}
\bibinfo{author}{\bibfnamefont{H.}~\bibnamefont{Haake}},
  \bibinfo{author}{\bibfnamefont{M.}~\bibnamefont{Ku\'{s}}}, \bibnamefont{and}
  \bibinfo{author}{\bibfnamefont{R.}~\bibnamefont{Scharf}},
  \bibinfo{journal}{Z. Phys. B} \textbf{\bibinfo{volume}{65}},
  \bibinfo{pages}{381} (\bibinfo{year}{1987}).

\bibitem[{\citenamefont{Shukla}(2005)}]{GUE4_21}
\bibinfo{author}{\bibfnamefont{P.}~\bibnamefont{Shukla}}, \bibinfo{journal}{J.
  Phys. Condens. Matter} \textbf{\bibinfo{volume}{17}}, \bibinfo{pages}{1653}
  (\bibinfo{year}{2005}).

\bibitem[{\citenamefont{Dyson}(1962)}]{GUE4_22}
\bibinfo{author}{\bibfnamefont{F.~J.} \bibnamefont{Dyson}},
  \bibinfo{journal}{J. Math. Phys.} \textbf{\bibinfo{volume}{3}},
  \bibinfo{pages}{1191} (\bibinfo{year}{1962}).

\bibitem[{\citenamefont{Stoffregen et~al.}(1995)\citenamefont{Stoffregen,
  Stein, St\"{o}ckmann, Ku\'{s}, and Haake}}]{QSC_26}
\bibinfo{author}{\bibfnamefont{U.}~\bibnamefont{Stoffregen}},
  \bibinfo{author}{\bibfnamefont{J.}~\bibnamefont{Stein}},
  \bibinfo{author}{\bibfnamefont{H.-J.} \bibnamefont{St\"{o}ckmann}},
  \bibinfo{author}{\bibfnamefont{M.}~\bibnamefont{Ku\'{s}}}, \bibnamefont{and}
  \bibinfo{author}{\bibfnamefont{F.}~\bibnamefont{Haake}},
  \bibinfo{journal}{Phys. Rev. Lett.} \textbf{\bibinfo{volume}{74}},
  \bibinfo{pages}{2666} (\bibinfo{year}{1995}).

\bibitem[{\citenamefont{Ponomarenko et~al.}(2008)\citenamefont{Ponomarenko,
  Schedin, Katsnelson, Yang, Hill, Novoselov, and Geim}}]{QC_9}
\bibinfo{author}{\bibfnamefont{L.~A.} \bibnamefont{Ponomarenko}},
  \bibinfo{author}{\bibfnamefont{F.}~\bibnamefont{Schedin}},
  \bibinfo{author}{\bibfnamefont{M.~I.} \bibnamefont{Katsnelson}},
  \bibinfo{author}{\bibfnamefont{R.}~\bibnamefont{Yang}},
  \bibinfo{author}{\bibfnamefont{E.~W.} \bibnamefont{Hill}},
  \bibinfo{author}{\bibfnamefont{K.~S.} \bibnamefont{Novoselov}},
  \bibnamefont{and} \bibinfo{author}{\bibfnamefont{A.~K.} \bibnamefont{Geim}},
  \bibinfo{journal}{Science} \textbf{\bibinfo{volume}{320}},
  \bibinfo{pages}{356} (\bibinfo{year}{2008}).

\bibitem[{\citenamefont{Chung et~al.}(2000)\citenamefont{Chung, Gokirmak, Wu,
  Bridgewater, Ott, Antonsen, and Anlage}}]{GUE7}
\bibinfo{author}{\bibfnamefont{S.-H.} \bibnamefont{Chung}},
  \bibinfo{author}{\bibfnamefont{A.}~\bibnamefont{Gokirmak}},
  \bibinfo{author}{\bibfnamefont{D.-H.} \bibnamefont{Wu}},
  \bibinfo{author}{\bibfnamefont{J.~S.~A.} \bibnamefont{Bridgewater}},
  \bibinfo{author}{\bibfnamefont{E.}~\bibnamefont{Ott}},
  \bibinfo{author}{\bibfnamefont{T.~M.} \bibnamefont{Antonsen}},
  \bibnamefont{and} \bibinfo{author}{\bibfnamefont{S.~M.}
  \bibnamefont{Anlage}}, \bibinfo{journal}{Phys. Rev. Lett.}
  \textbf{\bibinfo{volume}{85}}, \bibinfo{pages}{2482} (\bibinfo{year}{2000}).

\bibitem[{\citenamefont{Kazimierczuk et~al.}(2014)\citenamefont{Kazimierczuk,
  Fr\"{o}hlich, Scheel, Stolz, and Bayer}}]{GRE}
\bibinfo{author}{\bibfnamefont{T.}~\bibnamefont{Kazimierczuk}},
  \bibinfo{author}{\bibfnamefont{D.}~\bibnamefont{Fr\"{o}hlich}},
  \bibinfo{author}{\bibfnamefont{S.}~\bibnamefont{Scheel}},
  \bibinfo{author}{\bibfnamefont{H.}~\bibnamefont{Stolz}}, \bibnamefont{and}
  \bibinfo{author}{\bibfnamefont{M.}~\bibnamefont{Bayer}},
  \bibinfo{journal}{Nature} \textbf{\bibinfo{volume}{514}},
  \bibinfo{pages}{343} (\bibinfo{year}{2014}).

\bibitem[{\citenamefont{A{\ss}mann et~al.}(2016)\citenamefont{A{\ss}mann,
  Thewes, Fr\"{o}hlich, and Bayer}}]{QC}
\bibinfo{author}{\bibfnamefont{M.}~\bibnamefont{A{\ss}mann}},
  \bibinfo{author}{\bibfnamefont{J.}~\bibnamefont{Thewes}},
  \bibinfo{author}{\bibfnamefont{D.}~\bibnamefont{Fr\"{o}hlich}},
  \bibnamefont{and} \bibinfo{author}{\bibfnamefont{M.}~\bibnamefont{Bayer}},
  \bibinfo{journal}{Nature Mater.} \textbf{\bibinfo{volume}{15}},
  \bibinfo{pages}{741} (\bibinfo{year}{2016}).

\bibitem[{\citenamefont{Freitag et~al.}(2017)\citenamefont{Freitag,
  Heck\"{o}tter, Bayer, and A{\ss}mann}}]{QC2}
\bibinfo{author}{\bibfnamefont{M.}~\bibnamefont{Freitag}},
  \bibinfo{author}{\bibfnamefont{J.}~\bibnamefont{Heck\"{o}tter}},
  \bibinfo{author}{\bibfnamefont{M.}~\bibnamefont{Bayer}}, \bibnamefont{and}
  \bibinfo{author}{\bibfnamefont{M.}~\bibnamefont{A{\ss}mann}},
  \bibinfo{journal}{Phys. Rev. B} \textbf{\bibinfo{volume}{95}},
  \bibinfo{pages}{155204} (\bibinfo{year}{2017}).

\bibitem[{\citenamefont{Schweiner
  et~al.}(2017{\natexlab{a}})\citenamefont{Schweiner, Main, and Wunner}}]{175}
\bibinfo{author}{\bibfnamefont{F.}~\bibnamefont{Schweiner}},
  \bibinfo{author}{\bibfnamefont{J.}~\bibnamefont{Main}}, \bibnamefont{and}
  \bibinfo{author}{\bibfnamefont{G.}~\bibnamefont{Wunner}},
  \bibinfo{journal}{Phys. Rev. Lett.} \textbf{\bibinfo{volume}{118}},
  \bibinfo{pages}{046401} (\bibinfo{year}{2017}{\natexlab{a}}).

\bibitem[{\citenamefont{Schweiner
  et~al.}(2016{\natexlab{a}})\citenamefont{Schweiner, Main, and Wunner}}]{75}
\bibinfo{author}{\bibfnamefont{F.}~\bibnamefont{Schweiner}},
  \bibinfo{author}{\bibfnamefont{J.}~\bibnamefont{Main}}, \bibnamefont{and}
  \bibinfo{author}{\bibfnamefont{G.}~\bibnamefont{Wunner}},
  \bibinfo{journal}{Phys. Rev. B} \textbf{\bibinfo{volume}{93}},
  \bibinfo{pages}{085203} (\bibinfo{year}{2016}{\natexlab{a}}).

\bibitem[{\citenamefont{Gr\"unwald et~al.}(2016)\citenamefont{Gr\"unwald,
  A{\ss}mann, Heck\"{o}tter, Fr\"{o}hlich, Bayer, Stolz, and Scheel}}]{76}
\bibinfo{author}{\bibfnamefont{P.}~\bibnamefont{Gr\"unwald}},
  \bibinfo{author}{\bibfnamefont{M.}~\bibnamefont{A{\ss}mann}},
  \bibinfo{author}{\bibfnamefont{J.}~\bibnamefont{Heck\"{o}tter}},
  \bibinfo{author}{\bibfnamefont{D.}~\bibnamefont{Fr\"{o}hlich}},
  \bibinfo{author}{\bibfnamefont{M.}~\bibnamefont{Bayer}},
  \bibinfo{author}{\bibfnamefont{H.}~\bibnamefont{Stolz}}, \bibnamefont{and}
  \bibinfo{author}{\bibfnamefont{S.}~\bibnamefont{Scheel}},
  \bibinfo{journal}{Phys. Rev. Lett.} \textbf{\bibinfo{volume}{117}},
  \bibinfo{pages}{133003} (\bibinfo{year}{2016}).

\bibitem[{\citenamefont{Feldmaier et~al.}(2016)\citenamefont{Feldmaier, Main,
  Schweiner, Cartarius, and Wunner}}]{50}
\bibinfo{author}{\bibfnamefont{M.}~\bibnamefont{Feldmaier}},
  \bibinfo{author}{\bibfnamefont{J.}~\bibnamefont{Main}},
  \bibinfo{author}{\bibfnamefont{F.}~\bibnamefont{Schweiner}},
  \bibinfo{author}{\bibfnamefont{H.}~\bibnamefont{Cartarius}},
  \bibnamefont{and} \bibinfo{author}{\bibfnamefont{G.}~\bibnamefont{Wunner}},
  \bibinfo{journal}{J. Phys. B: At. Mol. Opt. Phys.}
  \textbf{\bibinfo{volume}{49}}, \bibinfo{pages}{144002}
  (\bibinfo{year}{2016}).

\bibitem[{\citenamefont{Thewes et~al.}(2015)\citenamefont{Thewes,
  Heck\"{o}tter, Kazimierczuk, A{\ss}mann, Fr\"{o}hlich, Bayer, Semina, and
  Glazov}}]{28}
\bibinfo{author}{\bibfnamefont{J.}~\bibnamefont{Thewes}},
  \bibinfo{author}{\bibfnamefont{J.}~\bibnamefont{Heck\"{o}tter}},
  \bibinfo{author}{\bibfnamefont{T.}~\bibnamefont{Kazimierczuk}},
  \bibinfo{author}{\bibfnamefont{M.}~\bibnamefont{A{\ss}mann}},
  \bibinfo{author}{\bibfnamefont{D.}~\bibnamefont{Fr\"{o}hlich}},
  \bibinfo{author}{\bibfnamefont{M.}~\bibnamefont{Bayer}},
  \bibinfo{author}{\bibfnamefont{M.~A.} \bibnamefont{Semina}},
  \bibnamefont{and} \bibinfo{author}{\bibfnamefont{M.~M.}
  \bibnamefont{Glazov}}, \bibinfo{journal}{Phys. Rev. Lett.}
  \textbf{\bibinfo{volume}{115}}, \bibinfo{pages}{027402}
  (\bibinfo{year}{2015}), \bibinfo{note}{and Supplementary Material}.

\bibitem[{\citenamefont{Sch\"{o}ne et~al.}(2016)\citenamefont{Sch\"{o}ne,
  Kr\"{u}ger, Gr\"{u}nwald, Stolz, Scheel, A{\ss}mann, Heck\"{o}tter, Thewes,
  Fr\"{o}hlich, and Bayer}}]{80}
\bibinfo{author}{\bibfnamefont{F.}~\bibnamefont{Sch\"{o}ne}},
  \bibinfo{author}{\bibfnamefont{S.~O.} \bibnamefont{Kr\"{u}ger}},
  \bibinfo{author}{\bibfnamefont{P.}~\bibnamefont{Gr\"{u}nwald}},
  \bibinfo{author}{\bibfnamefont{H.}~\bibnamefont{Stolz}},
  \bibinfo{author}{\bibfnamefont{S.}~\bibnamefont{Scheel}},
  \bibinfo{author}{\bibfnamefont{M.}~\bibnamefont{A{\ss}mann}},
  \bibinfo{author}{\bibfnamefont{J.}~\bibnamefont{Heck\"{o}tter}},
  \bibinfo{author}{\bibfnamefont{J.}~\bibnamefont{Thewes}},
  \bibinfo{author}{\bibfnamefont{D.}~\bibnamefont{Fr\"{o}hlich}},
  \bibnamefont{and} \bibinfo{author}{\bibfnamefont{M.}~\bibnamefont{Bayer}},
  \bibinfo{journal}{Phys. Rev. B} \textbf{\bibinfo{volume}{93}},
  \bibinfo{pages}{075203} (\bibinfo{year}{2016}).

\bibitem[{\citenamefont{Schweiner
  et~al.}(2016{\natexlab{b}})\citenamefont{Schweiner, Main, Feldmaier, Wunner,
  and Uihlein}}]{100}
\bibinfo{author}{\bibfnamefont{F.}~\bibnamefont{Schweiner}},
  \bibinfo{author}{\bibfnamefont{J.}~\bibnamefont{Main}},
  \bibinfo{author}{\bibfnamefont{M.}~\bibnamefont{Feldmaier}},
  \bibinfo{author}{\bibfnamefont{G.}~\bibnamefont{Wunner}}, \bibnamefont{and}
  \bibinfo{author}{\bibfnamefont{{\relax Ch}.}~\bibnamefont{Uihlein}},
  \bibinfo{journal}{Phys. Rev. B} \textbf{\bibinfo{volume}{93}},
  \bibinfo{pages}{195203} (\bibinfo{year}{2016}{\natexlab{b}}).

\bibitem[{\citenamefont{Schweiner
  et~al.}(2017{\natexlab{b}})\citenamefont{Schweiner, Main, Wunner, Freitag,
  Heck\"{o}tter, Uihlein, A{\ss}mann, Fr\"{o}hlich, and Bayer}}]{125}
\bibinfo{author}{\bibfnamefont{F.}~\bibnamefont{Schweiner}},
  \bibinfo{author}{\bibfnamefont{J.}~\bibnamefont{Main}},
  \bibinfo{author}{\bibfnamefont{G.}~\bibnamefont{Wunner}},
  \bibinfo{author}{\bibfnamefont{M.}~\bibnamefont{Freitag}},
  \bibinfo{author}{\bibfnamefont{J.}~\bibnamefont{Heck\"{o}tter}},
  \bibinfo{author}{\bibfnamefont{{\relax Ch}.}~\bibnamefont{Uihlein}},
  \bibinfo{author}{\bibfnamefont{M.}~\bibnamefont{A{\ss}mann}},
  \bibinfo{author}{\bibfnamefont{D.}~\bibnamefont{Fr\"{o}hlich}},
  \bibnamefont{and} \bibinfo{author}{\bibfnamefont{M.}~\bibnamefont{Bayer}},
  \bibinfo{journal}{Phys. Rev. B} \textbf{\bibinfo{volume}{95}},
  \bibinfo{pages}{035202} (\bibinfo{year}{2017}{\natexlab{b}}).

\bibitem[{\citenamefont{Heck\"otter et~al.}(2017)\citenamefont{Heck\"otter,
  Freitag, Fr\"ohlich, A{\ss}mann, Bayer, Semina, and Glazov}}]{78}
\bibinfo{author}{\bibfnamefont{J.}~\bibnamefont{Heck\"otter}},
  \bibinfo{author}{\bibfnamefont{M.}~\bibnamefont{Freitag}},
  \bibinfo{author}{\bibfnamefont{D.}~\bibnamefont{Fr\"ohlich}},
  \bibinfo{author}{\bibfnamefont{M.}~\bibnamefont{A{\ss}mann}},
  \bibinfo{author}{\bibfnamefont{M.}~\bibnamefont{Bayer}},
  \bibinfo{author}{\bibfnamefont{M.~A.} \bibnamefont{Semina}},
  \bibnamefont{and} \bibinfo{author}{\bibfnamefont{M.~M.}
  \bibnamefont{Glazov}}, \bibinfo{journal}{Phys. Rev. B}
  \textbf{\bibinfo{volume}{95}}, \bibinfo{pages}{035210}
  (\bibinfo{year}{2017}).

\bibitem[{\citenamefont{Zieli\'{n}ska-Raczy\'{n}ska
  et~al.}(2017)\citenamefont{Zieli\'{n}ska-Raczy\'{n}ska, Ziemkiewicz, and
  Czajkowski}}]{79}
\bibinfo{author}{\bibfnamefont{S.}~\bibnamefont{Zieli\'{n}ska-Raczy\'{n}ska}},
  \bibinfo{author}{\bibfnamefont{D.}~\bibnamefont{Ziemkiewicz}},
  \bibnamefont{and}
  \bibinfo{author}{\bibfnamefont{G.}~\bibnamefont{Czajkowski}},
  \bibinfo{journal}{Phys. Rev. B} \textbf{\bibinfo{volume}{95}},
  \bibinfo{pages}{075204} (\bibinfo{year}{2017}).

\bibitem[{\citenamefont{Schweiner
  et~al.}(2016{\natexlab{c}})\citenamefont{Schweiner, Main, Wunner, and
  Uihlein}}]{150}
\bibinfo{author}{\bibfnamefont{F.}~\bibnamefont{Schweiner}},
  \bibinfo{author}{\bibfnamefont{J.}~\bibnamefont{Main}},
  \bibinfo{author}{\bibfnamefont{G.}~\bibnamefont{Wunner}}, \bibnamefont{and}
  \bibinfo{author}{\bibfnamefont{{\relax Ch}.}~\bibnamefont{Uihlein}},
  \bibinfo{journal}{Phys. Rev. B} \textbf{\bibinfo{volume}{94}},
  \bibinfo{pages}{115201} (\bibinfo{year}{2016}{\natexlab{c}}).

\bibitem[{\citenamefont{Zieli\'{n}ska-Raczy\'{n}ska
  et~al.}(2016{\natexlab{a}})\citenamefont{Zieli\'{n}ska-Raczy\'{n}ska,
  Czajkowski, and Ziemkiewicz}}]{74}
\bibinfo{author}{\bibfnamefont{S.}~\bibnamefont{Zieli\'{n}ska-Raczy\'{n}ska}},
  \bibinfo{author}{\bibfnamefont{G.}~\bibnamefont{Czajkowski}},
  \bibnamefont{and}
  \bibinfo{author}{\bibfnamefont{D.}~\bibnamefont{Ziemkiewicz}},
  \bibinfo{journal}{Phys. Rev. B} \textbf{\bibinfo{volume}{93}},
  \bibinfo{pages}{075206} (\bibinfo{year}{2016}{\natexlab{a}}).

\bibitem[{\citenamefont{Zieli\'{n}ska-Raczy\'{n}ska
  et~al.}(2016{\natexlab{b}})\citenamefont{Zieli\'{n}ska-Raczy\'{n}ska,
  Ziemkiewicz, and Czajkowski}}]{77}
\bibinfo{author}{\bibfnamefont{S.}~\bibnamefont{Zieli\'{n}ska-Raczy\'{n}ska}},
  \bibinfo{author}{\bibfnamefont{D.}~\bibnamefont{Ziemkiewicz}},
  \bibnamefont{and}
  \bibinfo{author}{\bibfnamefont{G.}~\bibnamefont{Czajkowski}},
  \bibinfo{journal}{Phys. Rev. B} \textbf{\bibinfo{volume}{94}},
  \bibinfo{pages}{045205} (\bibinfo{year}{2016}{\natexlab{b}}).

\bibitem[{\citenamefont{Schweiner
  et~al.}(2017{\natexlab{c}})\citenamefont{Schweiner, Main, Wunner, and
  Uihlein}}]{200}
\bibinfo{author}{\bibfnamefont{F.}~\bibnamefont{Schweiner}},
  \bibinfo{author}{\bibfnamefont{J.}~\bibnamefont{Main}},
  \bibinfo{author}{\bibfnamefont{G.}~\bibnamefont{Wunner}}, \bibnamefont{and}
  \bibinfo{author}{\bibfnamefont{{\relax Ch}.}~\bibnamefont{Uihlein}},
  \bibinfo{journal}{Phys. Rev. B} \textbf{\bibinfo{volume}{95}},
  \bibinfo{pages}{195201} (\bibinfo{year}{2017}{\natexlab{c}}).

\bibitem[{\citenamefont{Klingshirn}(2007)}]{SO}
\bibinfo{author}{\bibfnamefont{C.}~\bibnamefont{Klingshirn}},
  \emph{\bibinfo{title}{Semiconductor Optics}} (\bibinfo{publisher}{Springer},
  \bibinfo{address}{Berlin}, \bibinfo{year}{2007}), \bibinfo{edition}{3rd} ed.

\bibitem[{\citenamefont{R\"{o}ssler}(2009)}]{SST}
\bibinfo{author}{\bibfnamefont{U.}~\bibnamefont{R\"{o}ssler}},
  \emph{\bibinfo{title}{Solid State Theory}} (\bibinfo{publisher}{Springer},
  \bibinfo{address}{Berlin}, \bibinfo{year}{2009}), \bibinfo{edition}{2nd} ed.

\bibitem[{\citenamefont{Knox}(1963)}]{TOE}
\bibinfo{author}{\bibfnamefont{R.}~\bibnamefont{Knox}},
  \emph{\bibinfo{title}{Theory of excitons}}, vol.~\bibinfo{volume}{5} of
  \emph{\bibinfo{series}{Solid State Physics Supplement}}
  (\bibinfo{publisher}{Academic}, \bibinfo{address}{New York},
  \bibinfo{year}{1963}).

\bibitem[{\citenamefont{Uihlein et~al.}(1981)\citenamefont{Uihlein,
  Fr\"{o}hlich, and Kenklies}}]{7}
\bibinfo{author}{\bibfnamefont{{\relax Ch}.}~\bibnamefont{Uihlein}},
  \bibinfo{author}{\bibfnamefont{D.}~\bibnamefont{Fr\"{o}hlich}},
  \bibnamefont{and} \bibinfo{author}{\bibfnamefont{R.}~\bibnamefont{Kenklies}},
  \bibinfo{journal}{Phys. Rev. B} \textbf{\bibinfo{volume}{23}},
  \bibinfo{pages}{2731} (\bibinfo{year}{1981}).

\bibitem[{\citenamefont{Kavoulakis et~al.}(1997)\citenamefont{Kavoulakis,
  Chang, and Baym}}]{1}
\bibinfo{author}{\bibfnamefont{G.~M.} \bibnamefont{Kavoulakis}},
  \bibinfo{author}{\bibfnamefont{Y.-C.} \bibnamefont{Chang}}, \bibnamefont{and}
  \bibinfo{author}{\bibfnamefont{G.}~\bibnamefont{Baym}},
  \bibinfo{journal}{Phys. Rev. B} \textbf{\bibinfo{volume}{55}},
  \bibinfo{pages}{7593} (\bibinfo{year}{1997}).

\bibitem[{\citenamefont{Wintgen}(1987)}]{GUE5_23}
\bibinfo{author}{\bibfnamefont{D.}~\bibnamefont{Wintgen}},
  \bibinfo{journal}{Phys. Rev. Lett.} \textbf{\bibinfo{volume}{58}},
  \bibinfo{pages}{1589} (\bibinfo{year}{1987}).

\bibitem[{\citenamefont{Wigner}(1957)}]{GUE4_8}
\bibinfo{author}{\bibfnamefont{E.~P.} \bibnamefont{Wigner}}, in
  \emph{\bibinfo{booktitle}{Conference on neutron physics by time-of-flight}},
  edited by \bibinfo{editor}{\bibfnamefont{R.~C.} \bibnamefont{Block}},
  \bibinfo{editor}{\bibfnamefont{W.~M.} \bibnamefont{Good}},
  \bibinfo{editor}{\bibfnamefont{J.~A.} \bibnamefont{Harvey}},
  \bibinfo{editor}{\bibfnamefont{H.~W.} \bibnamefont{Schmitt}},
  \bibnamefont{and} \bibinfo{editor}{\bibfnamefont{G.~T.}
  \bibnamefont{Trammell}} (\bibinfo{publisher}{Union Carbide Nuclear Company},
  \bibinfo{address}{Oak Ridge, Tennessee}, \bibinfo{year}{1957}), vol.
  \bibinfo{volume}{2309} of \emph{\bibinfo{series}{Oak Ridge National
  Laboratory Report}}, pp. \bibinfo{pages}{59--70}.

\bibitem[{\citenamefont{Lipari and Altarelli}(1977)}]{17_17}
\bibinfo{author}{\bibfnamefont{N.~O.} \bibnamefont{Lipari}} \bibnamefont{and}
  \bibinfo{author}{\bibfnamefont{M.}~\bibnamefont{Altarelli}},
  \bibinfo{journal}{Phys. Rev. B} \textbf{\bibinfo{volume}{15}},
  \bibinfo{pages}{4883} (\bibinfo{year}{1977}).

\bibitem[{\citenamefont{Luttinger}(1956)}]{25}
\bibinfo{author}{\bibfnamefont{J.}~\bibnamefont{Luttinger}},
  \bibinfo{journal}{Phys. Rev.} \textbf{\bibinfo{volume}{102}},
  \bibinfo{pages}{1030} (\bibinfo{year}{1956}).

\bibitem[{\citenamefont{Baldereschi and Lipari}(1973)}]{7_11}
\bibinfo{author}{\bibfnamefont{A.}~\bibnamefont{Baldereschi}} \bibnamefont{and}
  \bibinfo{author}{\bibfnamefont{N.~O.} \bibnamefont{Lipari}},
  \bibinfo{journal}{Phys. Rev. B} \textbf{\bibinfo{volume}{8}},
  \bibinfo{pages}{2697} (\bibinfo{year}{1973}).

\bibitem[{\citenamefont{Schmelcher and Cederbaum}(1992)}]{90}
\bibinfo{author}{\bibfnamefont{P.}~\bibnamefont{Schmelcher}} \bibnamefont{and}
  \bibinfo{author}{\bibfnamefont{L.~S.} \bibnamefont{Cederbaum}},
  \bibinfo{journal}{Z. Phys. D} \textbf{\bibinfo{volume}{24}},
  \bibinfo{pages}{311} (\bibinfo{year}{1992}).

\bibitem[{\citenamefont{Schmelcher and Cederbaum}(1993)}]{91}
\bibinfo{author}{\bibfnamefont{P.}~\bibnamefont{Schmelcher}} \bibnamefont{and}
  \bibinfo{author}{\bibfnamefont{L.~S.} \bibnamefont{Cederbaum}},
  \bibinfo{journal}{Phys. Rev. A} \textbf{\bibinfo{volume}{47}},
  \bibinfo{pages}{2634} (\bibinfo{year}{1993}).

\bibitem[{\citenamefont{Altarelli and Lipari}(1973)}]{34}
\bibinfo{author}{\bibfnamefont{M.}~\bibnamefont{Altarelli}} \bibnamefont{and}
  \bibinfo{author}{\bibfnamefont{N.~O.} \bibnamefont{Lipari}},
  \bibinfo{journal}{Phys. Rev. B} \textbf{\bibinfo{volume}{7}},
  \bibinfo{pages}{3798} (\bibinfo{year}{1973}).

\bibitem[{\citenamefont{Altarelli and Lipari}(1974)}]{33}
\bibinfo{author}{\bibfnamefont{M.}~\bibnamefont{Altarelli}} \bibnamefont{and}
  \bibinfo{author}{\bibfnamefont{N.~O.} \bibnamefont{Lipari}},
  \bibinfo{journal}{Phys. Rev. B} \textbf{\bibinfo{volume}{9}},
  \bibinfo{pages}{1733} (\bibinfo{year}{1974}).

\bibitem[{\citenamefont{Chen et~al.}(1987)\citenamefont{Chen, Gil, Mathieu, and
  Lascaray}}]{39}
\bibinfo{author}{\bibfnamefont{Y.}~\bibnamefont{Chen}},
  \bibinfo{author}{\bibfnamefont{B.}~\bibnamefont{Gil}},
  \bibinfo{author}{\bibfnamefont{H.}~\bibnamefont{Mathieu}}, \bibnamefont{and}
  \bibinfo{author}{\bibfnamefont{J.~P.} \bibnamefont{Lascaray}},
  \bibinfo{journal}{Phys. Rev. B} \textbf{\bibinfo{volume}{36}},
  \bibinfo{pages}{1510} (\bibinfo{year}{1987}).

\bibitem[{\citenamefont{Broeckx}(1991)}]{44}
\bibinfo{author}{\bibfnamefont{J.}~\bibnamefont{Broeckx}},
  \bibinfo{journal}{Phys. Rev. B} \textbf{\bibinfo{volume}{43}},
  \bibinfo{pages}{9643} (\bibinfo{year}{1991}).

\bibitem[{\citenamefont{Edmonds}(1960)}]{ED}
\bibinfo{author}{\bibfnamefont{A.}~\bibnamefont{Edmonds}},
  \emph{\bibinfo{title}{Angular momentum in quantum mechanics}}
  (\bibinfo{publisher}{Princeton University Press},
  \bibinfo{address}{Princeton}, \bibinfo{year}{1960}).

\bibitem[{\citenamefont{Caprio et~al.}(2012)\citenamefont{Caprio, Maris, and
  Vary}}]{S1}
\bibinfo{author}{\bibfnamefont{M.~A.} \bibnamefont{Caprio}},
  \bibinfo{author}{\bibfnamefont{P.}~\bibnamefont{Maris}}, \bibnamefont{and}
  \bibinfo{author}{\bibfnamefont{J.~P.} \bibnamefont{Vary}},
  \bibinfo{journal}{Phys. Rev. C} \textbf{\bibinfo{volume}{86}},
  \bibinfo{pages}{034312} (\bibinfo{year}{2012}).

\bibitem[{\citenamefont{Anderson et~al.}(1999)\citenamefont{Anderson, Bai,
  Bischof, Blackford, Demmel, Dongarra, Croz, Greenbaum, Hammarling, McKenney
  et~al.}}]{Lapack}
\bibinfo{author}{\bibfnamefont{E.}~\bibnamefont{Anderson}},
  \bibinfo{author}{\bibfnamefont{Z.}~\bibnamefont{Bai}},
  \bibinfo{author}{\bibfnamefont{C.}~\bibnamefont{Bischof}},
  \bibinfo{author}{\bibfnamefont{S.}~\bibnamefont{Blackford}},
  \bibinfo{author}{\bibfnamefont{J.}~\bibnamefont{Demmel}},
  \bibinfo{author}{\bibfnamefont{J.}~\bibnamefont{Dongarra}},
  \bibinfo{author}{\bibfnamefont{J.~D.} \bibnamefont{Croz}},
  \bibinfo{author}{\bibfnamefont{A.}~\bibnamefont{Greenbaum}},
  \bibinfo{author}{\bibfnamefont{S.}~\bibnamefont{Hammarling}},
  \bibinfo{author}{\bibfnamefont{A.}~\bibnamefont{McKenney}},
  \bibnamefont{et~al.}, \emph{\bibinfo{title}{{LAPACK} Users' Guide}}
  (\bibinfo{publisher}{Society for Industrial and Applied Mathematics},
  \bibinfo{address}{Philadelphia, PA}, \bibinfo{year}{1999}),
  \bibinfo{edition}{3rd} ed.

\bibitem[{\citenamefont{Harada and Hasegawa}(1983)}]{MA_24}
\bibinfo{author}{\bibfnamefont{H.}~\bibnamefont{Harada}} \bibnamefont{and}
  \bibinfo{author}{\bibfnamefont{H.}~\bibnamefont{Hasegawa}},
  \bibinfo{journal}{J. Phys. A} \textbf{\bibinfo{volume}{16}},
  \bibinfo{pages}{L259} (\bibinfo{year}{1983}).

\bibitem[{\citenamefont{Hasegawa et~al.}(1983)\citenamefont{Hasegawa, Adachi,
  and Harada}}]{MA_56}
\bibinfo{author}{\bibfnamefont{H.}~\bibnamefont{Hasegawa}},
  \bibinfo{author}{\bibfnamefont{S.}~\bibnamefont{Adachi}}, \bibnamefont{and}
  \bibinfo{author}{\bibfnamefont{A.}~\bibnamefont{Harada}},
  \bibinfo{journal}{J. Phys. A: Math. Gen.} \textbf{\bibinfo{volume}{16}},
  \bibinfo{pages}{L503} (\bibinfo{year}{1983}).

\bibitem[{\citenamefont{Messiah}(1969)}]{Messiah2}
\bibinfo{author}{\bibfnamefont{A.}~\bibnamefont{Messiah}},
  \emph{\bibinfo{title}{Quantum Mechanics 2}}
  (\bibinfo{publisher}{North-Holland}, \bibinfo{address}{Amsterdam},
  \bibinfo{year}{1969}).

\bibitem[{\citenamefont{Brody et~al.}(1981{\natexlab{b}})\citenamefont{Brody,
  Flores, French, Mello, Pandey, and Wong}}]{QC_16}
\bibinfo{author}{\bibfnamefont{T.~A.} \bibnamefont{Brody}},
  \bibinfo{author}{\bibfnamefont{J.}~\bibnamefont{Flores}},
  \bibinfo{author}{\bibfnamefont{J.~B.} \bibnamefont{French}},
  \bibinfo{author}{\bibfnamefont{P.~A.} \bibnamefont{Mello}},
  \bibinfo{author}{\bibfnamefont{A.}~\bibnamefont{Pandey}}, \bibnamefont{and}
  \bibinfo{author}{\bibfnamefont{S.~S.~M.} \bibnamefont{Wong}},
  \bibinfo{journal}{Rev. Mod. Phys.} \textbf{\bibinfo{volume}{53}},
  \bibinfo{pages}{385} (\bibinfo{year}{1981}{\natexlab{b}}).

\bibitem[{\citenamefont{Keppeler}(2003)}]{SPS}
\bibinfo{author}{\bibfnamefont{S.}~\bibnamefont{Keppeler}},
  \emph{\bibinfo{title}{Spinning Particles - Semiclassics and Spectral
  Statistics}} (\bibinfo{publisher}{Springer}, \bibinfo{address}{Berlin},
  \bibinfo{year}{2003}).

\bibitem[{\citenamefont{Grosa et~al.}(2014)\citenamefont{Grosa, Legranda,
  Mortessagnea, Richalotb, and Selemanib}}]{GUE2}
\bibinfo{author}{\bibfnamefont{J.-B.} \bibnamefont{Grosa}},
  \bibinfo{author}{\bibfnamefont{O.}~\bibnamefont{Legranda}},
  \bibinfo{author}{\bibfnamefont{F.}~\bibnamefont{Mortessagnea}},
  \bibinfo{author}{\bibfnamefont{E.}~\bibnamefont{Richalotb}},
  \bibnamefont{and}
  \bibinfo{author}{\bibfnamefont{K.}~\bibnamefont{Selemanib}},
  \bibinfo{journal}{Wave Motion} \textbf{\bibinfo{volume}{51}},
  \bibinfo{pages}{664} (\bibinfo{year}{2014}).

\bibitem[{\citenamefont{Abramowitz and Stegun}(1964)}]{GUE4_35}
\bibinfo{author}{\bibfnamefont{M.}~\bibnamefont{Abramowitz}} \bibnamefont{and}
  \bibinfo{author}{\bibfnamefont{I.~A.} \bibnamefont{Stegun}},
  \emph{\bibinfo{title}{Handbook of Mathematical Functions}}
  (\bibinfo{publisher}{Dover Publ.}, \bibinfo{address}{New York},
  \bibinfo{year}{1964}).

\bibitem[{\citenamefont{Berry and Robnik}(1984)}]{GUE3_10a}
\bibinfo{author}{\bibfnamefont{M.~V.} \bibnamefont{Berry}} \bibnamefont{and}
  \bibinfo{author}{\bibfnamefont{M.}~\bibnamefont{Robnik}},
  \bibinfo{journal}{J. Phys. A} \textbf{\bibinfo{volume}{17}},
  \bibinfo{pages}{2413} (\bibinfo{year}{1984}).

\bibitem[{\citenamefont{Brody}(1973)}]{GUE3_10b}
\bibinfo{author}{\bibfnamefont{T.~A.} \bibnamefont{Brody}},
  \bibinfo{journal}{Lett. Nuovo Cimento} \textbf{\bibinfo{volume}{7}},
  \bibinfo{pages}{482} (\bibinfo{year}{1973}).

\bibitem[{\citenamefont{Caurier et~al.}(1990)\citenamefont{Caurier,
  Grammaticos, and Ramani}}]{GUE3_10c}
\bibinfo{author}{\bibfnamefont{E.}~\bibnamefont{Caurier}},
  \bibinfo{author}{\bibfnamefont{B.}~\bibnamefont{Grammaticos}},
  \bibnamefont{and} \bibinfo{author}{\bibfnamefont{A.}~\bibnamefont{Ramani}},
  \bibinfo{journal}{J. Phys. A} \textbf{\bibinfo{volume}{23}},
  \bibinfo{pages}{4903} (\bibinfo{year}{1990}).

\bibitem[{\citenamefont{Hasegawa et~al.}(1988)\citenamefont{Hasegawa, Mikeska,
  and Frahm}}]{GUE3_10d}
\bibinfo{author}{\bibfnamefont{H.}~\bibnamefont{Hasegawa}},
  \bibinfo{author}{\bibfnamefont{H.~J.} \bibnamefont{Mikeska}},
  \bibnamefont{and} \bibinfo{author}{\bibfnamefont{H.}~\bibnamefont{Frahm}},
  \bibinfo{journal}{Phys. Rev. A} \textbf{\bibinfo{volume}{38}},
  \bibinfo{pages}{395} (\bibinfo{year}{1988}).

\bibitem[{\citenamefont{Izrailev}(1990)}]{GUE3_10e}
\bibinfo{author}{\bibfnamefont{F.}~\bibnamefont{Izrailev}},
  \bibinfo{journal}{Phys. Rep.} \textbf{\bibinfo{volume}{5-6}},
  \bibinfo{pages}{299} (\bibinfo{year}{1990}).

\bibitem[{\citenamefont{Gradshteyn and Ryzhik}(2007)}]{TI}
\bibinfo{author}{\bibfnamefont{I.}~\bibnamefont{Gradshteyn}} \bibnamefont{and}
  \bibinfo{author}{\bibfnamefont{I.}~\bibnamefont{Ryzhik}},
  \emph{\bibinfo{title}{Tables of Integrals, Series, and Products}}
  (\bibinfo{publisher}{Academic Press}, \bibinfo{address}{Burlington, MA},
  \bibinfo{year}{2007}), \bibinfo{edition}{7th} ed.

\bibitem[{\citenamefont{Mohr and Taylor}(2005)}]{50_50}
\bibinfo{author}{\bibfnamefont{P.~J.} \bibnamefont{Mohr}} \bibnamefont{and}
  \bibinfo{author}{\bibfnamefont{B.~N.} \bibnamefont{Taylor}},
  \bibinfo{journal}{Rev. Mod. Phys.} \textbf{\bibinfo{volume}{77}},
  \bibinfo{pages}{1} (\bibinfo{year}{2005}).

\end{thebibliography}
\end{document}